\documentclass[]{JHEP3}
\usepackage{epsfig}

\def\be{\begin{equation}}
\def\ee{\end{equation}}
\def\bea{\begin{array}}
\def\eea{\end{array}}
\def\beqa{\begin{eqnarray}}
\def\eeqa{\end{eqnarray}}
\def\beqas{\begin{eqnarray*}}
\def\eeqas{\end{eqnarray*}}

\def\bp{\begin{picture}}
\def\ep{\end{picture}}
\def\bc{\begin{center}}
\def\ec{\end{center}}
\def\bfig{\begin{figure}}
\def\efig{\end{figure}}

\def\bit{\begin{itemize}}
\def\eit{\end{itemize}}
\def\nn{\nonumber}
\def\f{\frac}

\def\[{\left[}
\def\]{\right]}
\def\({\left(}
\def\){\right)}

\def\..{\left.}
\def\.{\right.}
\def\tl{\tilde}

\def\tm{\times}

\def\bh{\supseteq}
\def\da{\dagger}

\def\al{\alpha}

\def\ep{\epsilon}

\def\de{\delta}

\def\pr{\prime}

\title{Realistic Standard Model Fermion Mass Relations in
Generalized Minimal Supergravity (GmSUGRA)}
\author{Csaba Balazs$^1$, Tianjun Li$^{2,3}$, Dimitri V.
Nanopoulos$^{2,4,5}$,
Fei Wang$^1$ \\
$^1$ School of Physics, Monash University, Melbourne Victoria 3800,
Australia\\
$^2$ George P. and Cynthia W. Mitchell Institute for Fundamental
Physics, Texas A$\&$M University, College Station, TX
77843, USA \\
$^3$ Key Laboratory of Frontiers in Theoretical Physics, Institute
of Theoretical Physics, Chinese Academy of Sciences,
Beijing 100190, P. R. China \\
$^4$ Astroparticle Physics Group, Houston Advanced Research Center
(HARC),
Mitchell Campus, Woodlands, TX 77381, USA \\
$^5$ Academy of Athens, Division of Natural Sciences, 28
Panepistimiou Avenue, Athens 10679, Greece }

\abstract{Grand Unified Theories (GUTs) usually predict
wrong Standard Model (SM) fermion mass relation
$m_e/m_{\mu} = m_d/m_s$ toward low energies.
To solve this problem, we consider
the Generalized Minimal Supergravity (GmSUGRA) models, which are
GUTs with gravity mediated supersymmetry breaking and higher dimensional
operators.  Introducing non-renormalizable terms in the
super- and K\"ahler potentials,
we can obtain the correct  SM fermion mass relations in the
$SU(5)$ model with GUT Higgs fields in the ${\bf 24}$ and ${\bf 75}$
representations, and in the $SO(10)$ model.  In the latter case the gauge symmetry
is broken down to $SU(3)_C\times SU(2)_L \times SU(2)_R
\times U(1)_{B-L}$, to flipped $SU(5)\times U(1)_X$,
or to $SU(3)_C\times SU(2)_L \times U(1)_1 \times U(1)_2$.
Especially, for the first time we generate the realistic SM fermion
mass relation in GUTs by considering the high-dimensional
operators in the K\"ahler potential.
}

\preprint{ACT-02-11, MIFPA-11-05}

\begin{document}
\maketitle \indent
\newpage
\section{Introduction}

Supersymmetry naturally solves the gauge hierarchy problem in the
Standard Model (SM). The unification of the
$SU(3)_C\times SU(2)_L \times U(1)_Y$ gauge couplings
 in the supersymmetric SM (SSM)
at about $2\tm 10^{16}$ GeV~\cite{Ellis:1990zq} strongly suggests
the existence of a Grand Unified Theory (GUT). In addition,
supersymmetric GUTs, such as the $SU(5)$~\cite{Georgi:1974sy} and
$SO(10)$~\cite{so10} models, give us deep insights into the
problems of the SM such as charge quantization, the origin of many free parameters,
the SM fermion masses and mixings, and beyond.
Although supersymmetric GUTs are attractive,
it is challenging to test them at the Large Hadron Collider (LHC),
the future International Linear Collider (ILC), or other
experiments.

In the traditional SSMs, supersymmetry is broken in the
hidden sector, and supersymmetry breaking effects can be mediated
to the observable sector via gravity~\cite{mSUGRA}, gauge
interactions~\cite{Ellis:1984bm, gaugemediation}, the super-Weyl
anomaly~\cite{anomalymediation, UVI-AMSB, D-AMSB}, or other
mechanisms. Recently, considering GUTs with gravity mediated
supersymmetry breaking and higher dimensional
operators~\cite{Ellis:1984bm, Ellis:1985jn, Hill:1983xh,
Shafi:1983gz, Drees:1985bx,
Anderson:1999uia, Chamoun:2001in, Chakrabortty:2008zk, Martin:2009ad,
Bhattacharya:2009wv, Feldman:2009zc, Chamoun:2009nd,stefan,india}
and F-theory GUTs with $U(1)$ fluxes~\cite{Vafa:1996xn,
Donagi:2008ca, Beasley:2008dc, Beasley:2008kw, Donagi:2008kj,
Font:2008id, Jiang:2009zza, Blumenhagen:2008aw, Jiang:2009za,
Li:2009cy, Leontaris:2009wi, Li:2010mr},
two of us (LN) proposed the Generalized Minimal Supregravity
 (GmSUGRA) scenario and studied
the generic gaugino mass relations as well as defined
their indices~\cite{Li:2010xr}. We also generalized
gauge and anomaly mediated supersymmetry breaking, and discussed
the corresponding gaugino mass relations and their indices~\cite{Li:2010hi}.

It is well known that one of the great successes of GUTs is
the prediction of the equal Yukawa couplings at the GUT scale
for the bottom ($b$) quark and $\tau$ lepton~\cite{Buras:1977yy}, which yields
the correct mass ratio $m_b/m_{\tau} \sim 2.7$ at the low
energy if and only if there are only three
generations~\cite{Nanopoulos:1978hh, Nanopoulos:1982fc}.
Alas, it is also well known that GUTs with minimal Higgs content
predict the wrong SM fermion mass relation $m_e/m_{\mu} = m_d/m_s$,
which is invariant under the renormalization group equation
(RGE) running due to the small Yukawa couplings of the first
two generations. This problem can be solved via the
Georgi-Jarlskog mechanism~\cite{georgi2}
by introducing Higgs fields in
higher dimensional representations in $SU(5)$
models (For generalization for $SO(10)$ models, see Ref.~\cite{gn}.),
via the Ellis-Gaillard mechanism~\cite{ellis} by introducing
higher dimensional operators (For generalization in the supersymmetric
models with  mass generation for the first two families of the SM fermions,
see Ref.~\cite{Nanopoulos:1982zm}.),  or invoking supersymmetric loop
effects~\cite{murayama}. Based on our previous work on
SM fermion Yukawa couplings in GmSUGRA~\cite{fei},
we aim to generate the
correct SM fermion mass relations in the $SU(5)$ and $SO(10)$
models.

In this paper, we briefly review GUTs and consider the general
gravity mediated supersymmetry breaking.  With non-renormalizable
terms in the superpotential~\cite{ellis, Nanopoulos:1982zm} and K\"ahler potential,
we can obtain the correct SM fermion mass relations
$m_em_s/m_dm_\mu\simeq 1/10$ in the $SU(5)$ model with
GUT Higgs fields in the ${\bf 24}$ and ${\bf 75}$
representations, and in $SO(10)$ model where
the gauge symmetry is broken down to
$SU(3)_C\times SU(2)_L \times SU(2)_R \times U(1)_{B-L}$,
to the flipped $SU(5)\times U(1)_X$ symmetry~\cite{smbarr, dimitri, AEHN-0},
or to $SU(3)_C\times SU(2)_L \times U(1)_1 \times U(1)_2$.
Our approach can be considered as the
generalizations of the Georgi-Jarlskog
and  Ellis-Gaillard mechanisms.
However, we cannot get realistic SM fermion mass
relations in $SO(10)$ models where the gauge symmetry
is broken down to the Pati-Salam $SU(4)_C\times SU(2)_L \times
SU(2)_R$ or to the George-Glashow $SU(5)\times
U(1)'$ symmetry. In the traditional Pati-Salam
and George-Glashow $SU(5)$ models, we predict
$m_e/m_{\mu} = m_d/m_s$. We emphasize that we for the first
time use the high-dimensional operators in the
K\"ahler potentail to derive the realistic SM fermion mass relation
in GUTs.


This paper is organized as follows. In Section~2,
we briefly review four-dimensional GUTs.
In Section~3, we explain general gravity mediated
supersymmetry breaking. With
higher dimensional operators in the super-
and K\"ahler potential, we  study the
SM fermion mass relations in $SU(5)$-based models in Section~4.
We consider $SO(10)$ models with higher dimensional operators in the super-
and K\"ahler potential in Section~5 and Section~6, respectively.
Section~\ref{sec-6} contains our conclusion.

\section{A Brief Review of Grand Unified Theories}
\label{sect-0}

In this Section we explain our conventions. In supersymmetric SMs,
we denote the left-handed quark doublets, right-handed up-type
quarks, right-handed down-type quarks, left-handed lepton doublets,
right-handed neutrinos, and right-handed charged leptons as $Q_i$,
$U^c_i$, $D^c_i$, $L_i$, $N^c_i$, and $E^c_i$, respectively.
We denote one pair of Higgs doublets as $H_u$ and $H_d$, which give
masses to the up-type quarks/neutrinos and the down-type
quarks/charged leptons, respectively. Moreover, we define
$\tan\beta \equiv \langle H_u^0 \rangle / \langle H_d^0 \rangle$,
where $v_{u,d} \equiv \langle H_{u,d}^0 \rangle $
are the Higgs vacuum expectation values (VEVs).

First, we briefly review the $SU(5)$ model. We define the $U(1)_{Y}$
hypercharge generator in $SU(5)$ as follows
\begin{eqnarray}
T_{\rm U(1)_{Y}}={\rm diag} \left(-{1\over 3}, -{1\over 3}, -{1\over
3}, {1\over 2}, {1\over 2} \right)~.~\, \label{u1y}
\end{eqnarray}
Under the $SU(3)_C\times SU(2)_L \times U(1)_Y$ gauge symmetry, the
$SU(5)$ representations are decomposed as follows
\begin{eqnarray}
\mathbf{5} &=& \mathbf{(3, 1, -1/3)} \oplus \mathbf{(1, 2, 1/2)}~,~ \\
\mathbf{\overline{5}} &=&
\mathbf{(\overline{3}, 1, 1/3)} \oplus \mathbf{(1, 2, -1/2)}~,~ \\
\mathbf{10} &=& \mathbf{(3, 2, 1/6)} \oplus \mathbf{({\overline{3}},
1, -2/3)}
\oplus \mathbf{(1, 1, 1)}~,~ \\
\mathbf{\overline{10}} &=& \mathbf{(\overline{3}, 2, -1/6)} \oplus
\mathbf{(3, 1, 2/3)}
\oplus \mathbf{(1, 1, -1)}~,~ \\
\mathbf{24} &=& \mathbf{(8, 1, 0)} \oplus \mathbf{(1, 3, 0)} \oplus
\mathbf{(1, 1, 0)} \oplus \mathbf{(3, 2, -5/6)} \oplus
\mathbf{(\overline{3}, 2, 5/6)}~.~\,
\end{eqnarray}
There are three families of the SM fermions whose quantum numbers
under $SU(5)$ are
\begin{eqnarray}
F'_i=\mathbf{10},~ {\overline f}'_i={\mathbf{\bar 5}},~
N^c_i={\mathbf{1}}~,~ \label{SU(5)-smfermions}
\end{eqnarray}
where $i=1, 2, 3$ for three families. The SM particle assignments in
$F'_i$ and ${\bar f}'_i$ are
\begin{eqnarray}
F'_i=(Q_i, U^c_i, E^c_i)~,~{\overline f}'_i=(D^c_i, L_i)~.~
\label{SU(5)-smparticles}
\end{eqnarray}
To break the $SU(5)$ and electroweak gauge symmetries,
we introduce the adjoint Higgs and another pair of Higgs fields
whose quantum numbers under $SU(5)$ are
\begin{eqnarray}
\Phi'~=~ {\mathbf{24}}~,~~~ h'~=~{\mathbf{5}}~,~~~{\overline
h}'~=~{\mathbf{\bar {5}}}~,~\, \label{SU(5)-1-Higgse}
\end{eqnarray}
where $h'$ and ${\overline h}'$ contain the Higgs doublets $H_u$ and
$H_d$, respectively.

Next, we briefly review the flipped $SU(5)\times U(1)_{X}$
model~\cite{smbarr, dimitri, AEHN-0}. The gauge group $SU(5)\times
U(1)_{X}$ can be embedded into $SO(10)$. We define the generator
$U(1)_{Y'}$ in $SU(5)$ as follows
\begin{eqnarray}
T_{\rm U(1)_{Y'}}={\rm diag} \left(-{1\over 3}, -{1\over 3},
-{1\over 3}, {1\over 2}, {1\over 2} \right). \label{u1yp}
\end{eqnarray}
The hypercharge is given by
\begin{eqnarray}
Q_{Y} = {1\over 5} \left( Q_{X}-Q_{Y'} \right). \label{ycharge}
\end{eqnarray}
The quantum numbers of the three SM fermion families
under $SU(5)\times U(1)_{X}$ are
\begin{eqnarray}
F_i={\mathbf{(10, 1)}},~ {\bar f}_i={\mathbf{(\bar 5, -3)}},~ {\bar
l}_i={\mathbf{(1, 5)}}, \label{smfermions}
\end{eqnarray}
where $i=1, 2, 3$. The particle assignments for the SM fermions are
\begin{eqnarray}
F_i=(Q_i, D^c_i, N^c_i)~,~~{\overline f}_i=(U^c_i,
L_i)~,~~{\overline l}_i=E^c_i~.~ \label{smparticles}
\end{eqnarray}
To break the GUT and electroweak gauge symmetries, we introduce two
pairs of Higgs fields whose quantum numbers under $SU(5)\times
U(1)_X$ are
\begin{eqnarray}
H={\mathbf{(10, 1)}}~,~~{\overline{H}}={\mathbf{({\overline{10}},
-1)}}~,~~ h={\mathbf{(5, -2)}}~,~~{\overline h}={\mathbf{({\bar
{5}}, 2)}}~,~\, \label{Higgse1}
\end{eqnarray}
where $h$ and ${\overline h}$ contain the Higgs doublets $H_d$ and
$H_u$, respectively.
The flipped $SU(5)\times U(1)_X$ model can be embedded
into $SO(10)$. Under the $SU(5)\times U(1)_X$ gauge symmetry, the
$SO(10)$ representations are decomposed as follows
\begin{eqnarray}
\mathbf{10} &=& \mathbf{(5, -2)} \oplus
\mathbf{(\overline{5}, 2)} ~,~ \\
\mathbf{{16}} &=& \mathbf{(10, 1)} \oplus \mathbf{(\overline{5},
-3)} \oplus
\mathbf{(1, 5)} ~,~\\
\mathbf{45} &=& \mathbf{(24, 0)} \oplus \mathbf{ (1, 0)} \oplus
\mathbf{(10, -4)} \oplus \mathbf{(\overline{10}, 4)} ~.~\,
\end{eqnarray}

Finally, we briefly review the Pati-Salam model. The gauge group is
$SU(4)_C \times SU(2)_L \times SU(2)_R$ which can also be embedded
into $SO(10)$. The quantum numbers of the three SM fermion families
under $SU(4)_C \times SU(2)_L \times SU(2)_R$ are
\begin{eqnarray}
F^L_i={\mathbf{(4, 2, 1)}}~,~~ F^{Rc}_i={\mathbf{(\overline{4}, 1,
2)}}~,~\,
\end{eqnarray}
where $i=1, 2, 3$. The particle assignments for the SM
fermions are
\begin{eqnarray}
F^L_i=(Q_i, L_i)~,~~F^{Rc}_i=(U^c_i, D^c_i, E^c_i, N^c_i)~.~\,
\end{eqnarray}
To break the Pati-Salam and electroweak gauge symmetries, we
introduce one pair of Higgs fields and one bi-doublet Higgs field
whose quantum numbers under $SU(4)_C \times SU(2)_L \times SU(2)_R$
are
\begin{eqnarray}
\Phi={\mathbf{(4, 1,
2)}}~,~~{\overline{\Phi}}={\mathbf{({\overline{4}}, 1, 2)}}~,~~
H'={\mathbf{(1, 2, 2)}}~,~\, \label{Higgse1}
\end{eqnarray}
where $H'$ contains one pair of the Higgs doublets $H_d$ and $H_u$.
The Pati-Salam model can be embedded into $SO(10)$ as well. Under
the $SU(4)_C\times SU(2)_L\times SU(2)_R$ gauge symmetry, the $SO(10)$
representations are decomposed as follows
\begin{eqnarray}
\mathbf{10} &=& \mathbf{(6, 1, 1)} \oplus
\mathbf{(1, 2, 2)} ~,~ \\
\mathbf{{16}} &=& \mathbf{(4, 2, 1)}
\oplus \mathbf{(\overline{4}, 1, 2)} ~,~\\
\mathbf{45} &=& \mathbf{(15, 1, 1)} \oplus \mathbf{ (1, 3, 1)}
\oplus \mathbf{ (1, 1, 3)} \oplus \mathbf{(6, 2, 2)} ~.~\,
\end{eqnarray}

\section{General Gravity Mediated Supersymmetry Breaking}
\label{sec-1}

The supegravity scalar potential can be written
as follows~\cite{mSUGRA} \beqa {V}=e^G\[G^i(G^{-1})^j_iG_j-3\]+\f{1}{2}
{\rm Re}\[(f^{-1})_{ab}\hat{D}^a\hat{D}^b\] ~,~\eeqa where D-terms are \beqa
\hat{D}^a{\equiv}-G^i(T^a)_i^j\phi_j=-\phi^{j*}(T^a)_j^iG_i~,\eeqa
and the K\"ahler function $G$ as well as its derivatives and the metric
$G_i^j$ are \beqa
G&{\equiv}&{K}+\ln\(W\)+\ln\({W^*}\)~,\\
G^i&=&\f{\delta G}{\de \phi_i}~,~~G_i=\f{\de G}{\de
\phi_i^*}~,~~G_i^j=\f{\de^2 G}{\de\phi^*_i\de\phi_j}~,~\, \eeqa
where $K$ is the K\"ahler potential and $W$ is the superpotential.

Since the gaugino masses, supersymmetry breaking scalar masses and
trilinear soft terms have been studied previously~\cite{Li:2010xr},
we only consider the SM fermion mass relations in this
paper. We consider the following K\"ahler potential
\begin{eqnarray}
K &=& \phi_i^{\dagger} e^{2gV}\phi_i +
\f{b_{\Phi \phi_i}}{M_*}\phi_i^{\dagger}e^{2gV}\Phi \phi_i+
\f{b_{S \phi_i}}{M_*}S\phi_i^{\dagger}e^{2gV}\phi_i~,
\end{eqnarray}
and superpotential \beqa W &=& \f{1}{6} y^{ijk} \phi_i \phi_j \phi_k
+ {1\over 6} \al^{ijk}_{\Phi} {{\Phi}\over M_*} \phi_i \phi_j \phi_k
~,\, \eeqa
where $M_*$ is the fundamental scale, $\Phi$ is the GUT Higgs field,
and $S$ is a SM singlet
Higgs field.

After the scalar components of the chiral superfields $\Phi$ and $S$
acquire vacuum expectation values (VEVs), we get the general superpotential
and K\"ahler potential
\begin{eqnarray}
K &=& a_{0\phi_i}\phi_i^\da e^{2gV}\phi_i+\f{b_{\Phi \phi_i}}{M_*}\phi_i^\da \langle \Phi \rangle
e^{2gV}\phi_i~,
\end{eqnarray}
\beqa W &=& \f{1}{6} y^{ijk} \phi_i \phi_j \phi_k + {1\over 6}
\al^{ijk}_{\Phi} {{ \langle \Phi \rangle}\over M_*} \phi_i \phi_j \phi_k ~, \eeqa
where
\begin{eqnarray}
a_{0\phi_i} ~=~ 1+b_{S \phi_i}\f{ \langle S \rangle}{M_*}~.~
\end{eqnarray}
Because $S$ is a SM singlet, it can acquire a VEV close to
the fundamental scale $M_*$ . Thus, $\langle S \rangle/M_*$ can be close
to 1 in principle. In short, the realistic SM fermion mass relations can be produced via
these non-renormalization terms in the superpotential and
K\"ahler potential~\cite{ellis, Nanopoulos:1982zm}. In particular, for the first time
we obtain the correct SM fermion mass relation in GUTs
via the high-dimensional operators in the K\"ahler potential.

\section{ $SU(5)$ Models}
\label{sect-2}

With non-renormalizable terms in the super- and
K\"ahler potentials, we generate the suitable SM fermion
mass ratio $m_em_s/m_{\mu}m_d$ in the $SU(5)$ models.
Before discussing the details, we summarize the realistic
SM fermion mass relations at the GUT scale.
Using low energy electroweak data, an effective universal
supersymmetry breaking scale of $M_S=500~{\rm GeV}$, and two-loop RGE running
for the SM gauge couplings and Yukawa couplings, we obtain
the SM fermion mass ratios at the GUT scale for the down-type
quarks and charged leptons~\cite{ross}:
\beqa
\f{m_b}{m_\tau}\approx1~,~~\f{3m_s}{m_\mu}\approx 0.69~,~~\f{m_d}{3m_e}\approx 0.83~.~
\eeqa
Due to the small Yukawa couplings this leads to the following RGE running
invariant SM fermion mass relation for the first two generations
\beqa
\f{m_e}{m_\mu}\approx\f{1}{10.8}\f{m_d}{m_s}~.
\eeqa
For comparison, standard mass ratios at the GUT scale are~\cite{georgi2}
\beqa
3m_e \approx m_d~,~~m_\mu \approx 3m_s~,~~m_\tau \approx m_b~,
\eeqa
which gives the RGE running invariant SM fermion mass ratio
\beqa
\f{m_e}{m_\mu}\approx\f{1}{9}\f{m_d}{m_s}~.
\eeqa

\subsection{Non-Renormalizable Terms in the Superpotential}

In this subsection, we study new contributions to
the SM fermion Yukawa couplings from higher dimensional
operators in the superpotential. To obtain the possible
higher dimensional operators for the Yukawa couplings, we need to
consider the decompositions of the tensor products for the SM
fermion Yukawa coupling terms~\cite{Slansky:1981yr} \beqa {\bf
10}\otimes{\bf 10}\otimes{\bf 5}&=&({\bf \bar{5}} \oplus {\bf
\overline{45}}\oplus{\bf \overline{50}} )\otimes {\bf 5} \nonumber
\\
&=& ({\bf 1} \oplus {\bf 24} ) \oplus ({\bf 24}\oplus {\bf
75}\oplus{\bf 126}) \oplus
({\bf 75}\oplus {\bf 175'}) ~,~ \\
{\bf 10}\otimes{\bf \bar{5}}\otimes {\bf \bar{5}}&=&{\bf
10}\otimes({\bf \overline{10}\oplus \overline{15}})=({\bf 1 \oplus
24\oplus 75})\oplus({\bf 24 \oplus \overline{126} })~. \eeqa

 Because the Higgs fields in the
${\bf 126}$, ${\bf \overline{126}}$ and ${\bf 175'}$ do not have the
$SU(3)_C\times SU(2)_L$ singlets~\cite{Slansky:1981yr}, we do not
consider them in the following discussions.  Thus, we only consider
the Higgs fields in the ${\bf 24}$ and ${\bf 75}$ representations. \\

(A) Higgs Field in the ${\bf 24}$ Representation. \\

The VEVs of the Higgs field $\Phi_{\bf 24}$ in the
adjoint representation can be expressed as
the following $5\times 5$ and $10\times 10$ matrices
\beqa
\langle \Phi_{\bf 24} \rangle ~=~ v {\sqrt
{3\over 5}} {\rm diag} \left(-{1\over 3}, -{1\over 3}, -{1\over 3},
{1\over 2}, {1\over 2} \right)~,~\,
\label{HVEV-24A}
\eeqa
\beqa \langle \Phi_{\bf
24} \rangle ~=~v \sqrt{\f{3}{5}} {\rm diag}
(\underbrace{-\f{2}{3},\cdots,
-\f{2}{3}}_3,\underbrace{~\f{1}{6},\cdots,~\f{1}{6}}_{6},~1)~,
\label{HVEV-24B}
\eeqa
which are normalized to $c=1/2$ and $c=3/2$, respectively.

For the Higgs field $\Phi_{\bf 24}$ in the ${\bf 24}$
representation, we consider the following superpotential for the
additional contributions to the SM fermion Yukawa coupling terms
\beqa W & \supset &\f{1}{M_*} \left(h^{Ui}
\epsilon^{mnpql} (F'_i)_{mn} (F'_i)_{pq} (h')_k (\Phi_{\bf 24})_l^k
+ h^{\prime Ui} \epsilon^{mnpkl} (F'_i)_{mn} (F'_i)_{pq} (h')_k
(\Phi_{\bf 24})_l^q \right. \nonumber \\ && \left. + h^{DEi}
(F'_i)_{mn} (\overline{f}'_i \otimes \overline{h}')_{Sym}^{ml}
(\Phi_{\bf 24})_l^n + h^{\prime DEi} (F'_i)_{mn} (\overline{f}'_i
\otimes \overline{h}')_{Asym}^{ml} (\Phi_{\bf 24})_l^n \right)~,~\,\eeqa
where the subscripts $Sym$ and $Asym$ denote the symmetric and
anti-symmetric products of two ${\bf \bar{5}}$ representations.
After $\Phi_{\bf 24}$ acquires a VEV, we obtain the Yukawa coupling
terms in the superpotential \beqa W & \supset & {v\over {M_*}}
\sqrt{\f{3}{5}} \left( -2 h^{Ui} Q_i U_i^c H_u - h^{\prime Ui} Q_i
U_i^c H_u -{1\over 6} h^{\prime DEi} Q_i D^c_i H_d - h^{\prime DEi}
L_i E^c_i H_d \right.\nonumber \\ &&\left. +{5\over 6} h^{ DEi} Q_i
D^c_i H_d  \right)~.~ \, \eeqa

For simplicity, we assume that the masses of the first generation are
dominanted by non-renormalizable terms, while the masses
of the second generation are generated as in the usual GUTs. Then we have the
following Yukawa coupling terms for the first generation
\beqa {\cal
L}\supseteq -c_1 (\f{1}{6}Q_1D_1^c H_d+L_1E_1^cH_d)+\f{5}{6}c_2
Q_1D_1^c H_d~,
\eeqa
where $c_1\approx \sqrt{\f{3}{5}} h^{\prime DE}\f{v}{M_*}$, and
$c_2\approx\sqrt{\f{3}{5}} h^{DE}\f{v}{M_*}$. We choose $c_i v_{d}\sim
{\cal O}({\rm MeV})$ which is at the order of the electron and
down quark masses. After electroweak symmetry breaking, choosing
$c_2\approx 12 c_1$, we
can obtain the correct RGE running invariant SM fermion
mass ratio  at the GUT scale
\beqa
\f{m_e m_s}{m_\mu m_d}=\f{6c_1}{5c_2-c_1}\approx \f{1}{10}~.~
\eeqa

(B) Higgs Field in the ${\bf 75}$ Representation. \\

The VEV of the ${\bf 75}$ dimensional Higgs field $\Phi_{jl}^{[ik]}$
can be written as follows~\cite{Ellis:1985jn}
\beqa
\langle
\Phi^{[ik]}_{jl} \rangle ~=~\f{v}{2\sqrt{3}}
\[\Delta_{cj}^{[i}\Delta_{cl}^{k]}
+2\Delta_{wj}^{[i}\Delta_{wl}^{k]}
-\f{1}{2}\delta^{[i}_{j}\delta^{k]}_{l}\]~,
\eeqa
where
\beqa
\Delta_c~=~{\rm diag}(~1,~1,~1,~0,~0)~,~~ \Delta_w~=~{\rm
diag}(~0,~0,~0,~1,~1)~.
\eeqa
We consider the following
superpotential for the additional contributions to the
SM fermion Yukawa coupling terms
\beqa
W & \supset & \left(h^{Ui} \epsilon^{mnpjl}
(F'_i)_{mn} (F'_i)_{pq} (h')_k \Phi^{[qk]}_{jl} + h^{\prime Ui}
\epsilon^{jlpqk} (F'_i)_{mn} (F'_i)_{pq} (h')_k \Phi^{[mn]}_{jl}
\right. \nonumber \\ && \left. + h^{DEi} (F'_i)_{mn}
(\overline{f}'_i)^p (\overline{h}')^q \Phi^{[mn]}_{pq} \right) ~.~\,
\eeqa
After $\Phi^{[ik]}_{jl}$ acquires a VEV, we obtain the Yukawa
coupling terms in the superpotential
\beqa
W & \supset & {v\over
{M_*}} \f{1}{2 {\sqrt {3}}} \left( - h^{\prime DEi} Q_i D^c_i H_d +3
h^{\prime DEi} L_i E^c_i H_d \right)~.~ \,
\eeqa

Similarly to the Georgi-Jarlskog mechanism~\cite{georgi2},
we can get the realistic SM fermion mass relation.
After imposing some discrete symmetry,
we can generate the following superpotential
\beqa \label{mass-ellis}
W&\bh& \(h^{ DE}_{12}Q_1D_2^c H_d+h^{ DE}_{12}L_1E_2^cH_d+h^{
DE}_{12}Q_2D_1^c H_d+h^{ DE}_{12}L_2E_1^cH_d \)\nn\\&+&{v\over
{M_*}} \f{1}{2 {\sqrt {3}}}\left( - h^{\prime DE}_{22} Q_2 D^c_2 H_d
+3 h^{\prime DE}_{22} L_2 E^c_2 H_d \right)~.
\eeqa
For not too large $\tan\beta$ and $h^{\pr DE}\sim {\cal O}(1)$,
we have $ h^{\pr DE} v_{d} v /M_*\sim {\cal O}(10^2)~{\rm MeV}$. Thus,
we get the following mass matrices for $(e,~\mu)$ and $(d,~s)$ after electroweak
symmetry breaking
\beqa
\bea{cc}&\bea{cc}e&~\mu\eea\\
\bea{c}e\\ \mu\eea &\(\bea{cc} ~0& ~a\\ ~a&~3b\eea\)~,\eea~~~~~\bea{cc}&\bea{cc}d&~s\eea\\
\bea{c}d\\ s\eea &\(\bea{cc} ~0& ~a\\ ~a&~b\eea\)~.
\eea
\eeqa
Diagonalizing these matrices for $a\ll b$, we can get approximately the
RG invariant SM fermion mass ratio \beqa \f{m_e}{m_{\mu}} \approx
\f{1}{9}\f{m_d}{m_s}~. \eeqa

\subsection{Non-Renormalizable Terms in the K\"ahler Potential}

In this subsection, we study the new contributions to
the SM fermion Yukawa couplings arising from higher dimensional
operators in the K\"ahler potential.
The realistic SM fermion mass ratios can also be produced
by the non-minimal K\"ahler potentials.
   In order to construct gauge invariant higher dimensional operators, we
need the decompositions of the following tensor products
\beqa {\bf
\bar{5}}\otimes{\bf 5}&=&{\bf 1}\oplus {\bf 24}~,~\, \eeqa \beqa
{\bf \overline{10}}\otimes {\bf 10}={\bf 1}\oplus {\bf 24}\oplus
{\bf 75}~.
\eeqa
Thus, the adjoint Higgs field can give additional contributions to the
kinetic terms for both $F'_i$ and $\overline{f}'_i$, while the Higgs
field in the ${\bf 75}$ representation can only give an extra contribution
to the kinetic term of $F'_i$.

For the non-minimal K\"ahler potential, the kinetic terms relevant
to $e,\mu,d,s$ are
\beqa K &{\bh}&Z_{Q_i}Q_i^\da Q_i+Z_{L_i}L_i^\da
L_i+Z_{E_i^c}(E_i^c)^\da (E_i^c)+Z_{D_i^c}(D_i^c)^\da (D_i^c)~.
\eeqa
With  the simple SM fermion Yukawa
coupling terms for the charged leptons and down-type quarks
\beqa W=y^{DE}_i
{F}^\pr_i \bar{f}^\pr_i \bar{h}~,
\eeqa
we obtain their masses after electroweak gauge symmetry breaking
 \beqa
m_e^i=\f{m^i_{DE}}{\sqrt{Z_{L^i}Z_{E_i^c}}}~,~~m_d^i=\f{m^i_{DE}}{\sqrt{Z_{Q^i}Z_{D_i^c}}}~.
\label{KT-SMF}
 \eeqa
Here $m^i_{DE}=y^{DE}_i \langle H_d \rangle $ are universal for the down-type quarks
and charged leptons in each
generations. In this work, we assume that each normalization factor $Z_{\Phi}$ is positive.\\

(A) Higgs Field in the ${\bf 24}$ Representation. \\

The VEVs of the Higgs field $\Phi_{\bf 24}$ in the
adjoint representation are given in Eqs. (\ref{HVEV-24A}) and
(\ref{HVEV-24B}). Thus, we
obtain the following normalizations for the SM fermion kinetic terms
\beqa
Z_{Q_i}&=&a_0+\sqrt{\f{3}{5}}\f{1}{6}\ep_1^i ,\\
Z_{U_i}&=&a_0-\sqrt{\f{3}{5}}\f{2}{3}\ep_1^i,\\
Z_{E_i^c}&=&a_0+\sqrt{\f{3}{5}}\ep_1^i,\\
Z_{D_i^c}&=&a'_0+\sqrt{\f{3}{5}}\f{1}{3}\ep_1^{\prime i} ,\\
Z_{L_i}&=&a'_0-\sqrt{\f{3}{5}}\f{1}{2}\ep_1^{\prime i}, \eeqa
where
\beqa a_0=1+b_{S {\bf 10}}\f{\langle S \rangle }{M_*}~,~~~~\ep_1^i=
b_{\Phi {\bf 10}}^i\f{\langle \Phi_{\bf 24} \rangle }{M_*}~, \label{ep} \eeqa
\beqa a'_0=1+b_{S {\bf \overline{5}}}\f{\langle S \rangle }{M_*}~,~~~~\ep_1^{\prime i}=
b_{\Phi {\bf \overline{5}}}^i\f{\langle \Phi_{\bf 24} \rangle }{M_*}~, \label{ep} \eeqa
where $i$ is the family index.

Thus, we can obtain the correct SM fermion mass ratio
 \beqa
\f{m_e m_s}{m_\mu
m_d}=\sqrt{\f{(b_1+\f{1}{6})(b'_1+\f{1}{3})(b_2+1)(b'_2-\f{1}{2})}
{(b_1+1)(b'_1-\f{1}{2})(b_2+\f{1}{6})(b'_2+\f{1}{3})}}\approx\f{1}{10}~.
 \eeqa
Here we normalize
\beqa
{a_0}={b_i}\sqrt{\f{3}{5}}\ep_1^i~,~~~{a'_0}={b'_i}\sqrt{\f{3}{5}}\ep_1^{\prime i}~,~
\eeqa
with no summation on the family index $i$. For instance,
we can choose $b_1 \approx b_2$, $b'_1 \not= \f{1}{2}$,
while $b'_2\approx\f{1}{2}$. \\

(B) Higgs Field in the ${\bf 75}$ Representation. \\

  Next, we consider the Higgs field $\Phi_{kl}^{[ij]}$ in the ${\bf
75}$ representation. Because the Higgs fields $\Phi_{\bf 24}$ and
$\Phi_{kl}^{[ij]}$ belong to the decomposition of the tensor  product
 $ {\bf \overline{10}} \tm {\bf 10}$, their VEVs
must be orthogonal to each other. Thus, we obtain the VEV of
$\Phi_{kl}^{[ij]}$ in terms of the $10\times 10$ matrix \beqa
\langle \Phi_{kl}^{[ij]} \rangle ~=~\f{v}{2\sqrt{3}} {\rm diag}
\left(\underbrace{~1,\cdots,~1}_3,\underbrace{-1,\cdots,-1}_{6},
3\right)~.~ \, \eeqa

So we obtain the normalizations for the SM fermion kinetic terms
\beqa
Z_{Q_i}&=&a_0-\f{1}{2\sqrt{3}}\ep_3^i,\\
Z_{U_i}&=&a_0-\f{1}{2\sqrt{3}}\ep_3^i,\\
Z_{E_i^c}&=&a_0+\f{3}{2\sqrt{3}}\ep_3^i,\\
Z_{{D}_i^c}&=& Z_{{L}_i} ~=~a_0~,~ \label{SUV-75SM} \eeqa
where
\beqa
a_0=1+b_{S {\bf 10}}\f{\langle S \rangle }{M_*}~,~~~~\ep_3^i=
b_{\Phi {\bf 10}}^i\f{\langle \Phi_{\bf 75} \rangle }{M_*}~,
\label{ep} \eeqa
and $i$ denotes the family index.
The realistic SM fermion mass ratio emerges as
 \beqa
\f{m_e m_s}{m_\mu
m_d}=\sqrt{\f{(b_1-1)(b_2+3)}{(b_2-1)(b_1+3)}}\approx\f{1}{10}~.
 \eeqa
 Here we normalize \beqa a_0= b_i\f{1}{2\sqrt{3}}\ep_3^i~,\eeqa with no summation on the
family index $i$.
For example, we can choose $b_2 \not= 1$ while
$b_1\approx 1$.

\section{$SO(10)$ Models with Non-Renormalizable Superpotential Terms}
\label{sect-3}

In the $SO(10)$ model, the  gauge symmetry can be broken directly down
to the Pati-Salam $SU(4)_C\times SU(2)_L \times SU(2)_R$,
the $SU(3)_C\times SU(2)_L \times SU(2)_R\times U(1)_{B-L}$ symmetry, the
Geogi-Glashow $SU(5)\times U(1)'$, and the
flipped $SU(5)\times U(1)_X$ symmetry. For the last two cases,
the gauge symmetry can be further reduced to the
$SU(3)_C \times SU(2)_L \times U(1)_1 \times U(1)_2$ symmetry.
In the Pati-Salam models and Georgi-Glashow
$SU(5)\times U(1)'$ models without further gauge
symmetry breaking, the masses for the down-type quarks and charged leptons
are the same. Thus, we cannot obtain the correct SM fermion mass
relations when we break the $SO(10)$ gauge symmetry down to
the $SU(4)_C\times SU(2)_L \times SU(2)_R$ or $SU(5)\times U(1)'$
symmetries. To be concrete, we shall also study these two scenarios
in details.

  There are several kinds of the renormalizable Yukawa coupling terms
for the SM fermions in the $SO(10)$ models. For example, we can
introduce the Higgs fields in the ${\bf 120}$ or ${\bf 126}$
representation to obtain
additional contributions to the SM fermion Yukawa couplings. In this
paper, we only consider the simplest Higgs fields\footnote{ We use
two ${\bf 10}$ Higgs to avoid large $\tan\beta$. } $H_{\bf
10}^i~(i=1,2)$ in the $SO(10)$ fundamental representation.
 The renormalizable terms in superpotential give the tree-level mass
 relations
\beqa m_{d^i}=m_{e^i}~, ~~~m_{u^i}=m_{\nu^i}~,\eeqa
after the Higgs fields $H_{\bf 10}^i$ acquire VEVs.
Due to the arbitrariness in neutrino sector, we will not discuss the
mass ratios for $u^i$ and $\nu^i$ here. We only consider the
SM fermion mass ratio
$m_em_s/m_{\mu} m_d$.

There are several ways to improve such mass ratio. For example, one
can introduce additional higher representation Higgs fields to generalize
the Georgi-Jarlskog mechanism in $SU(5)$ models~\cite{georgi2}
and Georgi-Nanopoulos mechanism in the $SO(10)$ models~\cite{gn}.
In this work, we generate the realistic SM fermion mass ratio
in the GmSUGRA, {\it i.e.} in the simple $SO(10)$ model with
 higher dimensional operators in the super- and
K\"ahler potentials. In this Section, we discuss the effects of
non-renormalizable terms in the superpotential on the SM fermion
mass relations.

  To obtain the non-renormalizable contributions to the SM fermion Yukawa
coupling terms, we need to know the decompositions of the tensor
product ${\bf 16}\otimes{\bf 16}\otimes {\bf
10}$~\cite{Slansky:1981yr} \beqa
{\bf 16}\otimes{\bf 16}&=&{\bf 10\oplus 120\oplus 126}~,\\
{\bf 16}\otimes{\bf 16}\otimes {\bf 10}&=&{\bf ( 1 \oplus 45\oplus
54)\oplus ( 45\oplus 210\oplus 945 )\oplus(210 \oplus 1050 ) }~.~\,
\eeqa Because the ${\bf 945}$ and ${\bf 1050}$ representations do
not have $SU(5)\times U(1)$ or $SU(4)_C\times SU(2)_L \times
SU(2)_R$ singlets~\cite{Slansky:1981yr}, we only consider the Higgs
fields in the ${\bf 45}$, ${\bf 54}$ and ${\bf 210}$
representations.



\subsection{The Pati-Salam Model}

The $SO(10)$ gauge symmetry can be broken down to the Pati-Salam
$SU(4)_C\times SU(2)_L \times SU(2)_R$ symmetry by giving VEVs
to the Higgs fields in the ${\bf 54}$ and ${\bf 210}$
representations.

We can write the VEV of the Higgs field $\Phi_{\bf 54}$ as
\beqa \langle \Phi_{\bf 54} \rangle =\f{v}{2\sqrt{15}} {\rm
diag}(\underbrace{~2,\cdots,~2}_6,\underbrace{-3,\cdots,-3}_4) ~,~\,
\eeqa which is normalized to $c=1$.

To calculate the additional contributions to the Yukawa coupling
terms, we consider the following superpotential \beqa W & \supset&
{1\over {M_*}} h^i ({\bf 16_i\otimes 16_i})_{\bf 10}^m (\Phi_{\bf
54})_{mn}{\bf 10}^n~.~\, \eeqa After $\Phi_{\bf 54}$ acquires a VEV,
we obtain the additional contributions to the SM fermion
Yukawa coupling terms
\beqa W & \supset& - h^i \f{3
v}{\sqrt{15}M_*}\left[{Q}_i{U}_i^cH_u+{L}_i{N}^c_i H_u + {Q}_iD_i^c
H_d + {L}_i{E_i^c} H_d \right]~.~\, \eeqa
Thus, the extra contributions to all the SM fermion Yukawa couplings
are the same, and then we cannot explain the SM fermion
mass ratio.

The VEV of the $\Phi_{\bf 210}$ Higgs field can be written as
\beqa \langle \Phi_{\bf 210} \rangle =\f{v}{2\sqrt{2}} {\rm
diag}(\underbrace{~1,\cdots,~1}_8,\underbrace{-1,\cdots,-1}_8) ~,~\,
\eeqa which is normalized to $c=2$. We consider the following
superpotential \beqa W & \supset& {1\over {M_*}} \left[ h^i ({\bf
16_i\otimes 16_i})_{\bf 120}^{mnl} (\Phi_{\bf 210})_{mnlk}{\bf 10}^k
+ h^{\prime i} ({\bf 16_i\otimes 16_i})_{\bf 126}^{mnlpq} (\Phi_{\bf
210})_{mnlp}{\bf 10}_q \right] ~.~\, \eeqa
It is easy to show that the above superpotential
will not contribute to the SM fermion Yukawa coupling terms.

In short, we cannot obtain the realistic SM fermion mass relation
since the Pati-Salam gauge symmetry is not broken. This problem
can be solved by
introducing additional renormalizable Yukawa coupling terms involving
the higher representation Higgs fields.

\subsection{The $SU(3)_C\tm SU(2)_L\tm SU(2)_R\tm U(1)_{B-L}$ Model}

The $SO(10)$ gauge symmetry can also be broken down to the $SU(3)_C\tm
SU(2)_L\tm SU(2)_R\tm U(1)_{B-L}$ gauge symmetry by giving VEVs to the $({\bf
15,~1,~1})$ components of the Higgs fields in the ${\bf 45}$ and
${\bf 210}$ representations under $SU(4)_C\times SU(2)_L \times
SU(2)_R$.

For the Higgs field $\Phi_{\bf 45}$ in the ${\bf 45}$
representation, the VEV can be written as
\beqa \langle \Phi_{\bf 45} \rangle =\f{v}{2\sqrt{6}}
{\rm diag}
(\underbrace{~2,\cdots,~2}_3,\underbrace{-2,\cdots,-2}_3,\underbrace{~0,\cdots,~0}_4)~,
\eeqa which is normalized as $c=1$.

To calculate the additional contributions to the SM fermion Yukawa coupling
terms, we consider the following
superpotential \beqa W & \supset& {1\over {M_*}} \left[ h^i ({\bf
16_i\otimes 16_i})_{\bf 10}^{m} (\Phi_{\bf 45})_{mn}{\bf 10}^n +
h^{\prime i} ({\bf 16_i\otimes 16_i})_{\bf 120}^{mnl} (\Phi_{\bf
45})_{mn}{\bf 10}_l \right]~.~\, \eeqa
However, the above superpotential will not contribute to the SM fermion
Yukawa coupling terms.

For the Higgs field $\Phi_{\bf 210}$ in the ${\bf 210}$
representation, the VEV is
\beqa \langle \Phi_{\bf 210} \rangle ~=~\f{v}{2\sqrt{6}}
{\rm diag}(\underbrace{~1,~1,~1,-3}_4)~, \eeqa with normalization
$c=2$. We consider the following superpotential \beqa W &
\supset& {1\over {M_*}} \left[ h^i ({\bf 16_i\otimes 16_i})_{\bf
120}^{mnl} (\Phi_{\bf 210})_{mnlk}{\bf 10}^k + h^{\prime i} ({\bf
16_i\otimes 16_i})_{\bf 126}^{mnlpq} (\Phi_{\bf 210})_{mnlp}{\bf
10}_q \right]~.~\, \eeqa After $\Phi_{\bf 210}$ acquires a VEV, we
obtain the additional contributions to the SM
fermion Yukawa coupling terms
\beqa W & \supset& h^{\prime i} \f{
v}{\sqrt{6}M_*}\left[{Q}_i{U}_i^cH_u - 3{L}_i{N}^c_i H_u +
{Q}_iD_i^c H_d -3 {L}_i{E_i^c} H_d \right]~.~\, \eeqa

Similar to the Georgi-Jarlskog mechanism
in $SU(5)$ models~\cite{georgi2} and Georgi-Nanopoulos mechanism in
$SO(10)$ models~\cite{gn}, we can explain the SM fermion mass ratio.
After imposing some discrete
symmetries, we can generate the following superpotential
\beqa W&\bh&
\(h^{ DE}_{12}Q_1D_2^c H_d+h^{
DE}_{12}L_1E_2^cH_d+h^{ DE}_{12}Q_2D_1^c H_d+h^{
DE}_{12}L_2E_1^cH_d \)\nn\\&+&{v\over {M_*}} \f{1}{{\sqrt
{6}}}\left( h^{\prime DE}_{22} Q_2 D^c_2 H_d -3 h^{\prime DE}_{22}
L_2 E^c_2 H_d \right). \eeqa
Again, with not too large $\tan\beta$ and $h^{\pr DE}\sim {\cal O}(1)$,
we have $ h^{\pr DE}_{22} v_{d} v /M_*\sim {\cal O}(10^2)~{\rm MeV}$.
Thus, we get the mass matrices for $(e,~\mu)$ and $(d,~s)$ after electroweak
symmetry breaking\beqa
\bea{cc}&\bea{cc}e&~\mu\eea\\
\bea{c}e\\ \mu\eea &\(\bea{cc} ~0& ~a\\ ~a&~3b\eea\)~,\eea~~~~~\bea{cc}&\bea{cc}d&~s\eea\\
\bea{c}d\\ s\eea &\(\bea{cc} ~0& ~a\\ ~a&~b\eea\)~.\eea \eeqa
Diagonalizing the mass matrices for $a\ll b$, we can get approximately the
RGE running invarian SM fermion mass ratio \beqa \f{m_e}{m_{\mu}}\sim
\f{1}{9}\f{m_d}{m_s}~. \eeqa




\subsection{The Georgi-Glashow $SU(5)\tm U(1)'$ Model }

The $SO(10)$ gauge symmetry can be broken down to the Georgi-Glashow
$SU(5)\tm U(1)'$ symmetry by giving VEVs to the Higgs fields
in the ${\bf 45}$ and ${\bf 210}$ representations.

For the Higgs field $\Phi_{\bf 45}$ in the ${\bf 45}$
representation, we can write the VEV in terms of the ${\bf 10\tm 10}$ matrix
\beqa \langle \Phi_{\bf 45} \rangle ~=~\f{v}{\sqrt{10}} {\rm
diag}(\underbrace{~1,\cdots,~1}_5,\underbrace{-1,\cdots,-1}_{5})~,~\,
\eeqa where the normalization is $c=1$. Using the conventions
in~\cite{He:1990jw} we obtain the non-zero components \beqa
(\Phi_{\bf 45})_{12}=(\Phi_{\bf 45})_{34}=(\Phi_{\bf 45})_{56}
=(\Phi_{\bf 45})_{78}=(\Phi_{\bf 45})_{90}= \f{v}{\sqrt{10}}~. \eeqa
To calculate the additional contributions to the
SM fermion Yukawa couplings,
we consider the following superpotential \beqa W & \supset&
{1\over {M_*}} \left[ h^i ({\bf 16_i\otimes 16_i})_{\bf 10}^m
(\Phi_{\bf 45})_{mn}{\bf 10}^n + h^{\prime i} ({\bf 16_i\otimes
16_i})_{\bf 120}^{mnl}(\Phi_{\bf 45})_{mn}{\bf 10}_l \right]~.~\,
\eeqa Note that ${\bf 120}$ is anti-symmetric representation, the
$h^{\prime i}$ term will not contribute to the SM fermion Yukawa
couplings. After $\Phi_{\bf 45}$ acquires a VEV, we obtain the
additional contributions to the Yukawa couplings \beqa W &
\supset& h^i \f{2 v}{\sqrt{10}M_*}\left[{Q}_i{U}_i^cH_u+{L}_i{N}^c_i
H_u -{Q}_iD_i^c H_d -{L}_i{E_i^c} H_d \right]~.~\, \eeqa
These terms
are the same for the down-type quarks and charged leptons, so
we cannot realize the correct SM fermion mass ratio.

For the Higgs field $\Phi_{\bf 210}$ in the ${\bf 210}$
representation, we can write the VEV in terms of the ${\bf 16\tm
16}$ matrix as follows \beqa \langle \Phi_{\bf 210} \rangle
~=~\f{v}{2\sqrt{5}} {\rm
diag}(\underbrace{~1,\cdots,~1}_5,\underbrace{-1,\cdots,-1}_{10},5)~,
\eeqa where the normalization is $c=2$. This VEV can be written in
components as follows \beqa (\Phi_{\bf 210})_{1234}&=& (\Phi_{\bf
210})_{1256}=(\Phi_{\bf 210})_{1278}= (\Phi_{\bf 210})_{1290}=
(\Phi_{\bf 210})_{3456} =(\Phi_{\bf 210})_{3478} \nn\\&=& (\Phi_{\bf
210})_{3490} =(\Phi_{\bf 210})_{5678}=(\Phi_{\bf 210})_{5690}
=(\Phi_{\bf 210})_{7890}=-\f{v}{2\sqrt{5}}~.~\, \eeqa We consider
the following superpotential \beqa W & \supset& {1\over {M_*}}
\left[ h^i ({\bf 16_i\otimes 16_i})_{\bf 120}^{mnl} (\Phi_{\bf
210})_{mnlk}{\bf 10}^k + h^{\prime i} ({\bf 16_i\otimes 16_i})_{\bf
126}^{mnlkp}(\Phi_{\bf 210})_{mnlk}{\bf 10}_p \right].~\, \eeqa
After $\Phi_{\bf 210}$ acquires a VEV, we obtain the additional
contributions to the SM fermion Yukawa couplings \beqa W & \supset&
h^{\prime i} \f{ v}{\sqrt{5}M_*}\left[3 {L}_i{N}^c_i H_u -{Q}_iU_i^c
H_u \right]~.~\, \eeqa

In summary, we cannot obtain the realistic SM fermion mass relations
in this case since the $SU(5)$ gauge symmetry is not broken. This problem
can be solved by introducing additional renormalizable Yukawa
coupling terms involving the higher representation Higgs fields.

\subsection{The Flipped $SU(5)\times U(1)_X$ Model}

The discussion for the flipped $SU(5)\times U(1)_X$ model is similar
to that of the Georgi-Glashow $SU(5)\times U(1)'$ model except
that we make the following transformations \beqa Q_i \leftrightarrow
Q_i~,~~U^c_i \leftrightarrow D_i^c~,~~ L_i \leftrightarrow
L_i~,~~N^c_i \leftrightarrow E_i^c~,~~ H_d \leftrightarrow H_u~.~\,
\label{GG-FSU5} \eeqa

Therefore, for the Higgs field in the ${\bf 45}$ representation, we
obtain the additional contributions to the SM fermion Yukawa
couplins \beqa W & \supset& h^i \f{2 v}{\sqrt{10}M_*}\left[
{Q}_iD_i^c H_d + {L}_i{E_i^c} H_d - {Q}_i{U}_i^cH_u - {L}_i{N}^c_i
H_u \right]~.~\, \eeqa
These contributions are the same
for the down-type quarks and charged leptons,
we cannot realize the correct SM fermion mass ratio.

For the Higgs field in the ${\bf 210}$
representation, we have \beqa W & \supset& h^{\prime i}
\f{v}{\sqrt{5}M_*}\left[3 {L}_i{E}^c_i H_d -{Q}_iD_i^c H_d
\right]~.~\, \eeqa
Similarly to the Georgi-Jarlskog and Georgi-Nanopoulos mechanisms or to our previous
discussion, we can generate the following
correct SM fermion mass ratio \beqa
\f{m_e}{m_{\mu}}\sim \f{1}{9}\f{m_d}{m_s}~. \eeqa




\subsection{The $SU(3)_C\tm SU(2)_L\tm U(1)_1\tm U(1)_{2}$ Model}

The $SO(10)$ gauge symmetry can be broken down to the $SU(3)_C\tm
SU(2)_L\tm U(1)_1\tm U(1)_{2}$ symmetry by giving VEVs to the
${\bf (24, 0)}$ component of the Higgs fields in the ${\bf 45}$,
${\bf 54}$ and ${\bf 210}$ representations under $SU(5)\times U(1)$,
or to the ${\bf (75, 0)}$ component of the Higgs field in the ${\bf
210}$ representation. In this subsection, we will study the SM
fermion Yukawa couplings in the $SO(10)$ model where
the gauge symmetry is broken down to the
$SU(3)_C\times SU(2)_L \times U(1)_Y \times U(1)'$
symmetry via the Georgi-Glashow $SU(5)\tm U(1)'$ symmetry.
We also comment on the SM fermion
 Yukawa couplings in the $SO(10)$ model where
the gauge symmetry is broken down to the
$SU(3)_C\times SU(2)_L \times U(1)_Y \times U(1)'$
symmetry via the flipped $SU(5)\times
U(1)_X$ symmetry, which can be obtained from the Georgi-Glashow
$SU(5)\tm U(1)'$ case by making the replacements in
Eq.~(\ref{GG-FSU5}).

First, for the Higgs field $\Phi_{\bf 45}$ in the ${\bf 45}$
representation, we can write the VEV in terms  of the ${\bf 10\tm
10}$ matrix as follows \beqa \langle \Phi_{\bf 45} \rangle
~=~v{\sqrt{\f{3}{5}}} {\rm
diag}(~\f{1}{3},~\f{1}{3},~\f{1}{3},-\f{1}{2},
-\f{1}{2},-\f{1}{3},-\f{1}{3},-\f{1}{3},~\f{1}{2},~\f{1}{2})~, \eeqa
which is normalized to $c=1$. It can also be written in components
as follows \beqa 3(\Phi_{\bf 45})_{12}=3(\Phi_{\bf 45})_{34}
=3(\Phi_{\bf 45})_{56}=-2 (\Phi_{\bf 45})_{78} =-2 (\Phi_{\bf
45})_{90}= v{\sqrt{\f{3}{5}}}~. \eeqa To calculate the additional
contributions to the SM fermion Yukawa couplings, we consider the
following superpotential \beqa W & \supset& {1\over {M_*}} \left[
h^i ({\bf 16_i\otimes 16_i})_{\bf 10}^m (\Phi_{\bf 45})_{mn}{\bf
10}^n + h^{\prime i} ({\bf 16_i\otimes 16_i})_{\bf
120}^{mnl}(\Phi_{\bf 45})_{mn}{\bf 10}_l \right]~.~\, \eeqa After
$\Phi_{\bf 45}$ acquires a VEV, we obtain additional
contributions to the SM fermion
Yukawa couplings \beqa W & \supset& h^i
\f{v}{2M_*} \sqrt{\f{3}{5}} \left[{Q}_i{U}_i^cH_u+{L}_i{N}^c_i H_u
-{Q}_iD_i^c H_d -{L}_i{E_i^c} H_d \right]~.~\, \eeqa
Since these terms are universal, we cannot obtain the correct
SM fermion mass ratio, and the same result holds for the intermediate
flipped $SU(5)\times U(1)_X$ model.

Second, for the Higgs field $\Phi_{\bf 54}$ in the ${\bf 54}$
representation, we can write the VEV in the ${\bf 10\tm
10}$ matrix form as follows \beqa \langle \Phi_{\bf 54} \rangle ~=~
v{\sqrt{\f{3}{5}}} {\rm diag}(~\f{1}{3},~\f{1}{3},~\f{1}{3},
-\f{1}{2},-\f{1}{2},~\f{1}{3},~\f{1}{3},~\f{1}{3},-\f{1}{2},-\f{1}{2})~,
\eeqa which is normalized to $c=1$. We consider the following
superpotential \beqa W & \supset& {1\over {M_*}} h^i ({\bf
16_i\otimes 16_i})_{\bf 10}^m (\Phi_{\bf 54})_{mn}{\bf 10}^n~.~\,
\eeqa After $\Phi_{\bf 54}$ acquires a VEV, we obtain the additional
contributions to the SM fermion Yukawa couplings \beqa W & \supset& - h^i
\f{v}{2M_*} \sqrt{\f{3}{5}} \left[{Q}_i{U}_i^cH_u+{L}_i{N}^c_i H_u +
{Q}_iD_i^c H_d + {L}_i{E_i^c} H_d \right]~.~\, \eeqa
Once again, we cannot get the realistic SM fermion mass ratio,
and the same result holds for the intermediate
flipped $SU(5)\times U(1)_X$ model.

Third, we consider that the ${\bf (24, 0)}$ component of the Higgs
field $\Phi^{\bf 24}_{\bf 210}$ in the ${\bf 210}$ representation
obtains a VEV. We can write its VEV in the ${\bf 16\tm 16}$ matrix
as follows \beqa \langle \Phi^{\bf 24}_{\bf 210} \rangle ~=~
\f{v}{\sqrt{5}} {\rm diag}(-1,-1,-1,~\f{3}{2},~\f{3}{2},
\underbrace{~\f{1}{6},\cdots,~\f{1}{6}}_6,-\f{2}{3},-\f{2}{3},-\f{2}{3},~1,~0)~,
\eeqa which is normalized to $c=2$. In components we have \beqa
6(\Phi^{\bf 24}_{\bf 210})_{1278} &=& 6(\Phi^{\bf 24}_{\bf
210})_{3478} =6(\Phi^{\bf 24}_{\bf 210})_{5678} =6(\Phi^{\bf
24}_{\bf 210})_{1290} \nonumber \\&=& 6(\Phi^{\bf 24}_{\bf
210})_{3490} =6(\Phi^{\bf 24}_{\bf 210})_{5690} =-\f{3}{2}(\Phi^{\bf
24}_{\bf 210})_{1234} \nonumber \\&=& -\f{3}{2}(\Phi^{\bf 24}_{\bf
210})_{1256} =-\f{3}{2} (\Phi^{\bf 24}_{\bf 210})_{3456} =
(\Phi^{\bf 24}_{\bf 210})_{7890}=\f{v}{\sqrt{5}} ~. \eeqa We
consider the following superpotential \beqa W & \supset& {1\over
{M_*}} \left[ h^i ({\bf 16_i\otimes 16_i})_{\bf 120}^{mnl}
(\Phi^{\bf 24}_{\bf 210})_{mnlk}{\bf 10}^k + h^{\prime i} ({\bf
16_i\otimes 16_i})_{\bf 126}^{mnlpq} (\Phi^{\bf 24}_{\bf
210})_{mnlp}{\bf 10}_q \right]~.~\, \eeqa After $\Phi^{\bf 24}_{\bf
210}$ acquires a VEV, the additional contributions to
the SM fermion
Yukawa couplings are \beqa W & \supset& h^{\prime i} \f{v}{M_*}
\f{1}{6\sqrt{5}} \left[-3 {Q}_i{U}_i^cH_u+9{L}_i{N}^c_i H_u -5
{Q}_iD_i^c H_d +15{L}_i{E_i^c} H_d \right]~.~\, \eeqa
Thus, similarly to the Georgi-Jarlskog mechanism, we can realize
the correct SM fermion mass ratio. The same result
holds for the intermediate flipped $SU(5)\times U(1)_X$ model.

Finally, we consider that the ${\bf (75, 0)}$ component of the Higgs
field $\Phi^{\bf 75}_{\bf 210}$ in the ${\bf 210}$ representation
obtains a VEV. We can write this VEV in the ${\bf 16\tm 16}$ matrix form
as follows \beqa \langle \Phi^{\bf 75}_{\bf 210} \rangle ~=~
\f{v}{3} {\rm
diag}(~0,~0,~0,~0,~0,\underbrace{-1,\cdots,-1}_6,~1,~1,~1,~3,~0)~,
\eeqa which is normalized to $c=2$. In components we have \beqa
(\Phi^{\bf 75}_{\bf 210})_{1278} &=& (\Phi^{\bf 75}_{\bf
210})_{3478} =(\Phi^{\bf 75}_{\bf 210})_{5678} =(\Phi^{\bf 75}_{\bf
210})_{1290} \nonumber \\&=& (\Phi^{\bf 75}_{\bf 210})_{3490}
=(\Phi^{\bf 75}_{\bf 210})_{5690} =-(\Phi^{\bf 75}_{\bf 210})_{1234}
\nonumber \\&=& -(\Phi^{\bf 75}_{\bf 210})_{1256} =- (\Phi^{\bf
75}_{\bf 210})_{3456} =- {1\over 3} (\Phi^{\bf 75}_{\bf
210})_{7890}= -\f{v}{3} ~. \eeqa We consider the following
superpotential \beqa W & \supset& {1\over {M_*}} \left[ h^i ({\bf
16_i\otimes 16_i})_{\bf 120}^{mnl} (\Phi^{\bf 75}_{\bf
210})_{mnlk}{\bf 10}^k + h^{\prime i} ({\bf 16_i\otimes 16_i})_{\bf
126}^{mnlpq} (\Phi^{\bf 75}_{\bf 210})_{mnlp}{\bf 10}_q \right]~.~\,
\eeqa After $\Phi^{\bf 75}_{\bf 210}$ acquires a VEV, we obtain the
additional contributions to the Yukawa couplings \beqa W &
\supset& h^{\prime i} \f{v}{3M_*} \left[- {Q}_iD_i^c H_d +
3{L}_i{E_i^c} H_d \right]~.~\, \eeqa
Again, similar to the Georgi-Jarlskog  and Georgi-Nanopoulos
mechanisms, we can obtain
the correct SM fermion mass ratio. However, in this case, we cannot
get the realistic SM fermion mass ratio in the intermediate
flipped $SU(5)\times U(1)_X$ model.

\section{$SO(10)$ Models with Non-Renormalizable Terms in the K\"ahler Potential}
\label{sec-4}
In this Section, we shall study the new contributions to the
SM fermion Yukawa couplings from higher dimensional operators in
the K\"ahler potential in the $SO(10)$ model.
Normalizing the Yukawa couplings \beqa W=\sum\limits_{ab,i=1}^2
y^{iDE}_{ab} ({\bf 16})^a({\bf 16})^b ({\bf 10}_i)~, \eeqa we obtain the
masses for the charged leptons and down-type quarks after electroweak
symmetry breaking, which are given in Eq. (\ref{KT-SMF}).

In order to
construct gauge invariant higher dimensional operators in the
K\"ahler potential, we need to decompose the tensor product of ${\bf
\overline{16}}\otimes{\bf 16}$ as follows \beqa {\bf
\overline{16}}\otimes{\bf 16}={\bf 1}\oplus{\bf 45}\oplus{\bf 210}
~.~\, \eeqa Thus, we only need to consider Higgs fields in the
${\bf 45}$ and ${\bf 210}$ representations. The $SO(10)$ gauge
symmetry can be broken down to the Pati-Salam $SU(4)_C \times
SU(2)_L \times SU(2)_R$ symmetry by the VEV of the
Higgs field in the
${\bf 210}$ representation, and can be
further broken to the $SU(3)_C\times SU(2)_L \times SU(2)_R
\times U(1)_{B-L}$ symmetry by the VEVs of the $({\bf
15}, {\bf 1}, {\bf 1})$ components of the Higgs fields in
the ${\bf 45}$ and ${\bf 210}$
representations under $SU(4)_C\times SU(2)_L
\times SU(2)_R$. In addition, the $SO(10)$ gauge symmetry can be
broken down to the Georgi-Glashow $SU(5)\times U(1)'$ and flipped
$SU(5)\times U(1)_X$ symmetries by Higgs fields in the
${\bf 45}$ and ${\bf 210}$ representations, and can be further
broken to the $SU(3)_C\times SU(2)_L \times U(1)_1 \times
U(1)_2$ gauge symmetries by the VEV of the $({\bf 24}, {\bf
0})$ component of the Higgs field in the ${\bf 45}$
representation under $SU(5)\times U(1)$, or by the VEVs of
the $({\bf 24}, {\bf 0})$ and $({\bf 75}, {\bf
0})$ components of the Higgs fields in the ${\bf 210}$ representation. Thus, in the
following, we consider all these gauge symmetry breaking chains.

\subsection{The Pati-Salam Model}

Decomposing the ${\bf \overline{16}}\otimes{\bf 16}$ tensor product of spinor
representations under the $SU(4)_C \times SU(2)_L \times SU(2)_R$
gauge symmetry, we obtain the VEV for the $({\bf 1,1,1})$ component
of the ${\bf 210}$ dimensional Higgs field $\Phi_{\bf 210}$ in terms of
the $16\tm 16$ matrix
\beqa \langle \Phi_{\bf 210} \rangle ~=~\f{v}{2\sqrt{2}} {\rm
diag} (\underbrace{~1,\cdots,~1}_8,\underbrace{-1,\cdots,-1}_8)
~,~\, \eeqa with the normalization $c=2$. This leads to the wave
function normalization of the SM fermions
\beqa Z_{{Q}_i}&=&a_0
+\f{1}{2\sqrt{2}}\beta_{\bf 210}^{\pr i}
\f{v}{M_*}~, \nonumber \\
Z_{{U}_i^c}&=&a_0 - \f{1}{2\sqrt{2}}\beta_{\bf 210}^{\pr i}
\f{v}{M_*} ~, \nonumber \\
Z_{{E}_i^c}&=&a_0 - \f{1}{2\sqrt{2}}\beta^{\pr i}_{\bf 210}
\f{v}{M_*} ~, \nonumber \\
Z_{{D}_i^c}&=&a_0 - \f{1}{2\sqrt{2}} \beta^{\pr i}_{\bf 210}
\f{v}{M_*}~, \nonumber \\
Z_{{L}_i}&=&a_0 + \f{1}{2\sqrt{2}} \beta^{\pr i}_{\bf 210}
\f{v}{M_*}~. \label{PSsm-210} \eeqa
From these, we cannot obtain the suitable SM fermion mass ratio.

\subsection{The $SU(3)_C\tm SU(2)_L\tm SU(2)_R\tm U(1)_{B-L}$ Model}

The $SO(10)$ gauge symmetry can be broken down to the $SU(3)_C\tm
SU(2)_L\tm SU(2)_R\tm U(1)_{B-L}$ symmetry by giving VEVs to the
$({\bf 15,1,1})$ components of the Higgs fields in the ${\bf 45}$ and
${\bf 210}$ representations under $SU(4)_C\times SU(2)_L \times
SU(2)_R$. The decomposition of ${\bf 16}$ under the $SU(3)_C\tm
SU(2)_L\tm SU(2)_R\tm U(1)_{B-L}$ symmetry is \beqa {\bf 16}={\bf
(3,2,1, {1/6})}\oplus {\bf (1,2,1, {-1/2})} \oplus {\bf
(\bar{3},1,\bar{2}, {-1/6})}\oplus {\bf (1,1,\bar{2}, {1/2})}~.
\eeqa

First, we consider the Higgs field $\Phi_{\bf 45}$ in the ${\bf
45}$ representation. The VEV of $\Phi_{\bf 45}$ can be written in
terms of the $16\tm 16$ matrix as follows \beqa
\langle \Phi_{\bf 45} \rangle ~=~\f{v}{2\sqrt{6}} {\rm
diag}(\underbrace{~1,~1,~1,-3}_2,\underbrace{-1,-1,-1,~3}_2)~, \eeqa
which is normalized as $c=2$. Then, the wave function normalization
for the SM fermions is
\beqa Z_{{Q}_i}&=&a_0+\f{1}{2\sqrt{6}} \beta^{\pr
i}_{\bf 45}\f{v}{M_*}~,
\nonumber \\
Z_{{U}_i^c}&=&a_0-\f{1}{2\sqrt{6}} \beta^{\pr i}_{\bf 45}
\f{v}{M_*} ~,\nonumber \\
Z_{{E}_i^c}&=&a_0+\f{3}{2\sqrt{6}} \beta^{\pr i}_{\bf 45}
\f{v}{M_*} ~,\nonumber \\
Z_{{D}_i^c}&=&a_0-\f{1}{2\sqrt{6}} \beta^{\pr i}_{\bf 45}
\f{v}{M_*}~,\nonumber \\
Z_{{L}_i}&=&a_0-\f{3}{2\sqrt{6}} \beta^{\pr i}_{\bf 45}
\f{v}{M_*}~. \eeqa
Thus, we can obtain the correct SM fermion mass ratio
 \beqa
\f{m_e m_s}{m_\mu
m_d}=\sqrt{\f{(b_1-1)(b_1+1)(b_2-3)(b_2+3)}{(b_1+3)(b_1-3)(b_2+1)(b_2-1)}}\approx\f{1}{10}~.
 \eeqa
Here we normalize
\beqa {a_0}={b_i}\f{1}{2\sqrt{6}} \beta^{\pr
i}_{\bf 45}\f{v }{M_*}~,
\eeqa with no summation on the family index
$i$. For example, we can choose $b_1\not = 3$ and $b_2\not = 1$ while
 $b_2\approx 3$.

Second, we consider the Higgs field $\Phi_{\bf 210}$ in the ${\bf
210}$ representation. The VEV of $\Phi_{\bf 210}$ in terms of a
$16\tm 16$ matrix is \beqa \langle \Phi_{\bf 210} \rangle ~=~
\f{v}{2\sqrt{6}} {\rm diag}(\underbrace{~1,~1,~1,-3}_4)~, \eeqa
which is normalized as $c=2$. Thus, the wave function normalization
for the SM fermions is
\beqa Z_{{Q}_i}&=&a_0+\f{1}{2\sqrt{6}} \beta'_{\bf 210}
\f{v}{M_*}~,
\nonumber \\
Z_{{U}_i^c}&=&a_0+\f{1}{2\sqrt{6}} \beta^{\pr i}_{\bf 210}
\f{v}{M_*} ~, \nonumber \\
Z_{{E}_i^c}&=&a_0-\f{3}{2\sqrt{6}} \beta^{\pr i}_{\bf 210}
\f{v}{M_*} ~, \nonumber \\
Z_{{D}_i^c}&=&a_0+\f{1}{2\sqrt{6}} \beta^{\pr i}_{\bf 210}
\f{v}{M_*}~,\nonumber \\
Z_{{L}_i}&=&a_0-\f{3}{2\sqrt{6}} \beta^{\pr i}_{\bf 210}
\f{v}{M_*}~. \label{SOT-210} \eeqa
So we can obtain the realistic SM fermion mass ratio
 \beqa
\f{m_e m_s}{m_\mu
m_d}=\sqrt{\f{(b_1+1)^2(b_2-3)^2}{(b_1-3)^2(b_2+1)^2}}\approx\f{1}{10}~.
 \eeqa
Here we normalize \beqa {a_0}={b_i}\f{1}{2\sqrt{6}} \beta^{\pr
i}_{\bf 210}\f{v }{M_*}~,\eeqa
with no summation on the family index
$i$. For instance, we can choose $b_1\not= 3$ while $b_2\approx3$.

\subsection{The Georgi-Glashow $SU(5)\tm U(1)'$ and Flipped $SU(5)\times
U(1)_X$ Models}

The $SO(10)$ gauge symmetry can also be broken down to the $SU(5)\tm
U(1)$ symmetry by the VEVs of the ${\bf 45}$ and ${\bf 210}$ dimensional
Higgs fields $\Phi_{\bf 45}$ and $\Phi_{\bf 210}$. The decomposition
of the ${\bf 16}$ spinor representation under $SU(5)\tm U(1)$ is
\beqa {\bf 16}={\bf (10, ~1)\oplus (\bar{5},~-3)\oplus (1,~5)} ~.
\label{SOT-GGFSUV} \eeqa

(A) Higgs Field in the $\mathbf{45}$ Representation. \\

First, we consider the Higgs field $\Phi_{\bf 45}$. From
Eq.~(\ref{SOT-GGFSUV}), we obtain the VEV of $\Phi_{\bf 45}$ in
terms of the $16\tm 16$ matrix \beqa \langle \Phi_{\bf 45} \rangle ~=~
\f{v}{2\sqrt{10}} {\rm
diag}(\underbrace{-3,\cdots,-3}_5,\underbrace{1,\cdots,1}_{10},5)~,
\eeqa which is normalized as $c=2$. Consequently, we obtain the wave
function normalization in the Georgi-Glashow $SU(5)\tm U(1)'$ and
flipped $SU(5)\times U(1)_X$ models:
\begin{itemize}
\item The Georgi-Glashow $SU(5)\tm U(1)'$ Model
\beqa Z({F}'_i)&=&a_0+\beta^{\pr i}_{\bf 45}\f{v }{2\sqrt{10}M_*}
~, \nonumber \\
Z({\overline{f}}'_i)&=&a_0-3 \beta^{\pr i}_{\bf 45}\f{v
}{2\sqrt{10}M_*}~, \nonumber \\
Z({N}^c_i)&=&a_0 + 5 \beta^{\pr i}_{\bf 45}\f{v }{2\sqrt{10}M_*}~.
\eeqa
We cannot obtain the correct SM fermion mass relation
in the symmetry breaking chain from $SO(10)$
down to the Georgi-Glashow $SU(5)\tm U(1)'$ gauge symmetry
since $SU(5)$ is not broken.

\item The Flipped $SU(5)\times U(1)_X$ Model
\beqa Z({F}_i)&=&a_0+\beta^{\pr i}_{\bf 45}\f{v }{2\sqrt{10}M_*}
~, \nonumber \\
Z({\overline{f}}_i)&=&a_0-3 \beta^{\pr i}_{\bf 45}\f{v
}{2\sqrt{10}M_*}~, \nonumber \\
Z({\overline{l}}_i)&=&a_0 + 5 \beta^{\pr i}_{\bf 45}\f{v
}{2\sqrt{10}M_*}~. \eeqa
\end{itemize}

In the symmetry breaking chain from $SO(10)$ to
the flipped $SU(5)\times U(1)_X$ gauge symmetry,
we can get the realistic SM fermion mass ratio
\beqa \f{m_e m_s}{m_\mu
m_d}=\sqrt{\f{(b_1+1)^2(b_2-3)(b_2+5)}{(b_2+1)^2(b_1-3)(b_1+5)}}\approx\f{1}{10}~.
 \eeqa
Here we normalize \beqa {a_0}={b_i}\beta^{\pr i}_{\bf 45}\f{v
}{2\sqrt{10}M_*}~,\eeqa with no summation on the family index $i$.
We can choose $b_1\not= 3$ while $b_2\approx3$. \\




(B) Higgs Field in the $\mathbf{210}$ Representation. \\


We consider the $\Phi_{\bf 210}$ Higgs field,
the VEV of which is orthogonal
to that of the $\Phi_{\bf 45}$
\beqa \langle \Phi \rangle =\f{v}{2\sqrt{5}} {\rm
diag}(\underbrace{~1,\cdots,~1}_5,\underbrace{-1,\cdots,-1}_{10},~5)~,
\eeqa and is normalized as $c=2$. So we obtain the wave
function normalizations for the SM fermions
in the Georgi-Glashow $SU(5)\tm U(1)'$ and
flipped $SU(5)\times U(1)_X$ models:
\begin{itemize}
\item The Georgi-Glashow $SU(5)\tm U(1)'$ Model
\beqa Z({F}'_i)&=&a_0-\beta^{\pr i}_{\bf 210}\f{v }{2\sqrt{5}M_*}
~, \nonumber \\
Z({\overline{f}}'_i)&=&a_0 + \beta^{\pr i}_{\bf 210}\f{v
}{2\sqrt{5}M_*}~, \nonumber \\
Z({N}^c_i)&=&a_0 + 5 \beta^{\pr i}_{\bf 210}\f{v}{2\sqrt{5}M_*}~.
\eeqa

Thus, we cannot obtain the suitable SM fermion mass relation
in the symmetry breaking chain from the $SO(10)$ gauge symmetry
down to the Georgi-Glashow $SU(5)\tm U(1)'$ gauge symmetry
since the $SU(5)$ gauge symmetry is not broken.

\item The Flipped $SU(5)\times U(1)_X$ Model
\beqa Z(\tl{F}_i)&=&a_0 - \beta^{\pr i}_{\bf 210}\f{v
}{2\sqrt{5}M_*}
~, \nonumber \\
Z(\tl{\overline{f}}_i)&=&a_0 + \beta^{\pr i}_{\bf 210}\f{v
}{2\sqrt{5}M_*}~, \nonumber \\
Z(\tl{\overline{l}}_i)&=&a_0 + 5 \beta^{\pr i}_{\bf 210}\f{v
}{2\sqrt{5}M_*}~. \eeqa
\end{itemize}

In the symmetry breaking chain from the $SO(10)$ gauge symmetry down
to the flipped
$SU(5) \times U(1)_X$ gauge symmetry, we can realize the
correct SM fermion mass ratio \beqa
\f{m_e m_s}{m_\mu
m_d}=\sqrt{\f{(b_1-1)^2(b_2+1)(b_2+5)}{(b_2-1)^2(b_1+1)(b_1+5)}}\approx\f{1}{10}~.
 \eeqa
Here we normalize \beqa {a_0}={b_i}\beta^{\pr i}_{\bf 210}\f{v
}{2\sqrt{5}M_*}~,\eeqa with no summation on the family index $i$.
For instance, we can choose $b_2\not= 1 $ while $b_1\approx1$.



\subsection{The $SU(3)_C\tm SU(2)_L\tm U(1)_1\tm U(1)_2$ Model}

The $SO(10)$ gauge symmetry can also be broken down to the $SU(3)_C\tm
SU(2)_L\tm U(1)_1 \tm U(1)_{2}$ symmetry by the VEV of the ${\bf (24, 0)}$
component of the Higgs field in the ${\bf 45}$ representation
under $SU(5)\times U(1)$, or by the VEVs of the ${\bf (24, 0)}$ and ${\bf (75,
0)}$ components of the Higgs fields in the ${\bf 210}$ representation. \\

(A) Higgs Field in  the ${\bf (24, 0)}$
Component of the $\mathbf{45}$ Representation. \\

First, we consider the Higgs field $\Phi^{\bf 24}_{\bf 45}$ in the
 ${\bf 45}$ representation whose  ${\bf (24, 0)}$ component  acquires
the following VEV
 \beqa \langle \Phi_{\bf 45}^{\bf 24}
\rangle ~=~v \sqrt{\f{3}{5}} {\rm
diag}(~\f{1}{3},~\f{1}{3},~\f{1}{3},-\f{1}{2},
-\f{1}{2},\underbrace{~\f{1}{6},\cdots,~\f{1}{6}}_6,-\f{2}{3},
-\f{2}{3},-\f{2}{3},~1,~0)~, \eeqa which is normalized to $c=2$.

From this, we obtain the wave function normalizations for the
SM fermions in the
Georgi-Glashow $SU(5)\tm U(1)'$ and flipped $SU(5)\times U(1)_X$
models:
\begin{itemize}
\item The Georgi-Glashow $SU(5)\tm U(1)'$ Model
\beqa Z_{{Q}_i}&=&a_0+\sqrt{\f{3}{5}} \beta^{\pr i{\bf 24}}_{\bf 45}
\f{1}{6}\f{v}{M_*}~, \nonumber \\
Z_{{U}_i^c}&=&a_0-\sqrt{\f{3}{5}} \beta^{\pr i{\bf 24}}_{\bf 45}
\f{2}{3}\f{v}{M_*} ~, \nonumber \\
Z_{{E}_i^c}&=&a_0+\sqrt{\f{3}{5}} \beta^{\pr i{\bf 24}}_{\bf
45}\f{v}{M_*}
~, \nonumber \\
Z_{{D}^c_i}&=&a_0+\sqrt{\f{3}{5}} \beta^{\pr i{\bf 24}}_{\bf 45}
\f{1}{3}\f{v}{M_*} ~, \nonumber \\
Z_{{L}_i}&=&a_0-\sqrt{\f{3}{5}} \beta^{\pr i{\bf 24}}_{\bf 45}
\f{1}{2}\f{v}{M_*} ~.~\, \eeqa

In the symmetry breaking chain from the $SO(10)$ gauge symmetry
via Georgi-Glashow $SU(5)\tm U(1)^\pr$ down
to the $SU(3)_C\tm SU(2)_L\tm U(1)_1\tm U(1)_2$ symmetry, we
can get the correct SM fermion mass ratio
\beqa \f{m_e m_s}{m_\mu
m_d}=\sqrt{\f{(b_1+\f{1}{3})(b_1+\f{1}{6})(b_2-\f{1}{2})(b_2+1)}
{(b_1-\f{1}{2})(b_1+1)(b_2+\f{1}{3})(b_2+\f{1}{6})}}\approx\f{1}{10}~.
 \eeqa
Here we normalize \beqa {a_0}={b_i}\beta^{\pr i{\bf 24}}_{\bf 45}
\sqrt{\f{3}{5}}\f{v}{M_*}~,\eeqa with no summation on the family
 index $i$. For instance, we can choose $b_1\not= \f{1}{2}$
while $b_2\approx\f{1}{2}$.

\item The Flipped $SU(5)\times U(1)_X$ Model
\beqa
 Z_{{Q}_i}&=&a_0+\sqrt{\f{3}{5}}\beta^{\pr i{\bf 24}}_{\bf 45}
\f{1}{6}\f{v}{M_*} ~, \nonumber \\
Z_{{U}_i^c}&=&a_0+\sqrt{\f{3}{5}} \beta^{\pr i{\bf 24}}_{\bf 45}
\f{1}{3}\f{v}{M_*} ~,\\
Z_{{E}_i^c}&=&a_0~, \nonumber \\
Z_{{D}^c_i}&=&a_0-\sqrt{\f{3}{5}} \beta^{\pr i{\bf 24}}_{\bf 45}
\f{2}{3}\f{v}{M_*} ~, \nonumber \\
Z_{{L}_i}&=&a_0-\sqrt{\f{3}{5}} \beta^{\pr i{\bf 24}}_{\bf 45}
\f{1}{2}\f{v}{M_*} ~. \eeqa
\end{itemize}

In the symmetry breaking chain from the $SO(10)$ gauge symmetry
via flipped $SU(5)\tm U(1)_X$ down
to the $SU(3)_C\tm SU(2)_L\tm U(1)_1\tm U(1)_2$ gauge symmetry,
 we can obtain the realistic SM fermion
mass ratio \beqa \f{m_e m_s}{m_\mu
m_d}=\sqrt{\f{(b_1-\f{2}{3})(b_1+\f{1}{6})(b_2-\f{1}{2})b_2}
{(b_1-\f{1}{2})b_1(b_2-\f{2}{3})(b_2+\f{1}{6})}}\approx\f{1}{10}~.
 \eeqa
Here we normalize \beqa {a_0}={b_i}\beta^{\pr i{\bf 24}}_{\bf 45}
\sqrt{\f{3}{5}}\f{v}{M_*}~,\eeqa with no summation on the family
index $i$. For example, we can choose $b_1\not= \f{1}{2}$ and
$b_2\not= \f{2}{3}$ while $b_1\approx\f{2}{3}$
and/or $b_2\approx\f{1}{2}$. \\

(B) Higgs Field in  the ${\bf (24, 0)}$
Component of the $\mathbf{210}$ Representation. \\

Second, we consider the Higgs field $\Phi^{\bf 24}_{\bf 210}$ in
 the ${\bf 210}$ representation whose ${\bf (24, 0)}$ component
acquires a VEV as follows \beqa \langle \Phi^{\bf 24}_{\bf 210}
\rangle ~=~\f{v}{\sqrt{5}}{\rm diag}(-1,-1,-1,~\f{3}{2},~\f{3}{2},
\underbrace{~\f{1}{6},\cdots,~\f{1}{6}}_6,-\f{2}{3},-\f{2}{3},-\f{2}{3},~1,~0)~,
\eeqa which is normalized to $c=2$. In this case the wave
function normalizations for the SM fermions
via the Georgi-Glashow $SU(5)\tm U(1)'$ and
the flipped $SU(5)\times U(1)_X$ models are:

\begin{itemize}
\item The Georgi-Glashow $SU(5)\tm U(1)'$ Model

\beqa Z_{{Q}_i}&=&a_0+\f{1}{\sqrt{5}} \beta^{\prime i{\bf 24}}_{\bf
210}
\f{1}{6}\f{v}{M_*} ~, \nonumber \\
Z_{{U}_i^c}&=&a_0-\f{1}{\sqrt{5}} \beta^{\prime i{\bf 24}}_{\bf 210}
\f{2}{3}\f{v}{M_*}~, \nonumber \\
Z_{{E}_i^c}&=&a_0+\f{1}{\sqrt{5}} \beta^{\prime i{\bf 24}}_{\bf 210}
\f{v}{M_*}~, \nonumber \\
Z_{{D}^c_i}&=&a_0-\f{1}{\sqrt{5}} \beta^{\prime i{\bf 24}}_{\bf 210}
\f{v}{M_*} ~, \nonumber \\
Z_{{L}_i}&=&a_0+\f{1}{\sqrt{5}} \beta^{\prime i{\bf 24}}_{\bf 210}
\f{3}{2} \f{v}{M_*}~. \eeqa

In the symmetry breaking chain from the $SO(10)$ gauge symmetry
via Georgi-Glashow $SU(5)\tm U(1)^\pr$ down
to the $SU(3)_C\tm SU(2)_L\tm U(1)_1\tm U(1)_2$ symmetry, we
can get the realistic SM fermion mass ratio \beqa \f{m_e m_s}{m_\mu
m_d}=\sqrt{\f{(b_1-1)(b_1+\f{1}{6})(b_2+\f{3}{2})(b_2+1)}
{(b_1+\f{3}{2})(b_1+1)(b_2+\f{1}{6})(b_2-1)}}\approx\f{1}{10}~.
 \eeqa
Here we normalize \beqa {a_0}={b_i}\beta^{\pr i{\bf 24}}_{\bf 210}
\f{1}{\sqrt{5}}\f{v}{M_*}~,\eeqa with no summation on the family
 index $i$. For example, we can choose $b_2\not=1$ while $b_1\approx1$.

\item The Flipped $SU(5)\times U(1)_X$ Model

\beqa Z_{{Q}_i}&=&a_0+\f{1}{\sqrt{5}} \beta^{\prime i{\bf 24}}_{\bf
210}
\f{1}{6}\f{v}{M_*} ~, \nonumber \\
Z_{{U}_i^c}&=&a_0-\f{1}{\sqrt{5}} \beta^{\prime i{\bf 24}}_{\bf 210}
\f{v}{M_*} ~, \nonumber \\
Z_{{E}_i^c}&=&a_0~, \nonumber \\
Z_{{D}^c_i}&=&a_0-\f{1}{\sqrt{5}} \beta^{\prime i{\bf 24}}_{\bf 210}
\f{2}{3}\f{v}{M_*} ~, \nonumber \\
Z_{{L}_i}&=&a_0+\f{1}{\sqrt{5}} \beta^{\prime i{\bf 24}}_{\bf 210}
\f{3}{2}\f{v}{M_*} ~. \eeqa
\end{itemize}

In the symmetry breaking chain from the $SO(10)$ gauge symmetry
via flipped $SU(5)\tm U(1)_X$
down to the  $SU(3)_C\tm SU(2)_L\tm U(1)_1\tm U(1)_2$ symmetry,
we can obtain
the correct SM fermion mass ratio \beqa \f{m_e m_s}{m_\mu
m_d}=\sqrt{\f{(b_1+\f{1}{6})(b_1-\f{2}{3})(b_2+\f{3}{2})b_2}{(b_1+\f{3}{2})b_1(b_2+\f{1}{6})(b_2-\f{2}{3})}}\approx\f{1}{10}~.
 \eeqa
Here we normalize \beqa {a_0}={b_i}\beta^{\pr i{\bf 24}}_{\bf 210}
\f{1}{\sqrt{5}}\f{v}{M_*}~,\eeqa with no summation on the family
 index $i$. For instance, we can choose $b_2\not= \f{2}{3}$
while $b_1\approx\f{2}{3}$. \\

(C) Higgs Field in  the ${\bf (75, 0)}$
Component of the $\mathbf{210}$ Representation. \\

Third, we consider the Higgs field $\Phi^{\bf 75}_{\bf 210}$ in
 the ${\bf 210}$ representation whose ${\bf (75, 0)}$ component
acquires the following VEV
\beqa \langle \Phi^{\bf 75}_{\bf 210} \rangle
~=~\f{v}{3}{\rm
diag}(~0,~0,~0,~0,~0,\underbrace{-1,\cdots,-1}_6,~1,~1,~1,~3,~0)~,~\,
\eeqa which is normalized to $c=2$. Thus, we obtain the
following wave function normalizations
for the SM fermions via the Georgi-Glashow
$SU(5)\tm U(1)'$ and the flipped $SU(5)\times U(1)_X$ models:

\begin{itemize}
\item The Georgi-Glashow $SU(5)\tm U(1)'$ Model

\beqa Z_{{Q}_i}&=&a_0-\f{1}{3} \beta^{\prime i{\bf 75}}_{\bf 210}
\f{v}{M_*}~, \nonumber \\
Z_{{U}_i^c}&=&a_0+\f{1}{3} \beta^{\prime i{\bf 75}}_{\bf 210}
\f{v}{M_*} ~, \nonumber \\
Z_{{E}_i^c}&=&a_0+ \beta^{\prime i{\bf 75}}_{\bf 210}
\f{v}{M_*}~, \nonumber \\
Z_{{D}^c_i}&=&a_0~, \nonumber \\
Z_{{L}_i}&=&a_0~. \eeqa

In the symmetry breaking chain from the $SO(10)$ gauge symmetry
via Georgi-Glashow $SU(5)\tm U(1)^\pr$ down
to the  $SU(3)_C\tm SU(2)_L\tm U(1)_1\tm U(1)_2$ symmetry, we
can obtain the correct SM fermion mass ratio \beqa \f{m_e m_s}{m_\mu
m_d}=\sqrt{\f{(b_1-1)(b_2+3)}{(b_1+3)(b_2-1)}}\approx\f{1}{10}~.
 \eeqa
Here we normalize \beqa {a_0}={b_i}\f{1}{3} \beta^{\prime i{\bf
75}}_{\bf 210} \f{v}{M_*}~,\eeqa with no summation on the family
 index $i$. For instance, we can choose $b_2\not= 1$ while $b_1\approx 1$.

\item The Flipped $SU(5)\times U(1)_X$ Model

\beqa Z_{{Q}_i}&=&a_0-\f{1}{3} \beta^{\prime i{\bf 75}}_{\bf 210}
\f{v}{M_*} ~, \nonumber \\
Z_{{U}_i^c}&=&a_0~, \nonumber \\
Z_{{E}_i^c}&=&a_0~, \nonumber \\
Z_{{D}^c_i}&=&a_0+\f{1}{3} \beta^{\prime i{\bf 75}}_{\bf 210}
\f{v}{M_*} ~, \nonumber \\
Z_{{L}_i}&=&a_0~. \eeqa
\end{itemize}




In the symmetry breaking chain from the $SO(10)$ gauge symmetry
via flipped $SU(5)\tm U(1)_X$ down
to the $SU(3)_C\tm SU(2)_L\tm U(1)_1\tm U(1)_2$ symmetry,
we can get the realistic SM fermion
mass ratio \beqa \f{m_e m_s}{m_\mu
m_d}=\sqrt{\f{(b_1-1)(b_1+1)b_2^2}{b_1^2(b_2-1)(b_2+1)}}\approx\f{1}{10}~.
 \eeqa
Here we normalize \beqa {a_0}={b_i}\f{1}{3} \beta^{\prime i{\bf
75}}_{\bf 210} \f{v}{M_*}~,\eeqa with no summation on the family
 index $i$. For instance, we can choose $b_2\not= 1$ while $b_1\approx 1$.

\section{Conclusion}
\label{sec-6}
   Grand Unified Theories (GUTs) usually predict
wrong Standard Model (SM) fermion mass relations, such as
$m_e/m_{\mu} = m_d/m_s$, toward low energies. Based on our previous work on the SM fermion Yukawa couplings
in the GmSUGRA scenario with the higher dimensional
operators containing the GUT Higgs fields, we studied the SM fermion mass relations.
Considering non-renormalizable terms in the
super- and K\"ahler potentials,
we can obtain the correct  SM fermion mass relations in the
$SU(5)$ model with GUT Higgs fields in the ${\bf 24}$ and ${\bf 75}$
representations, and in the $SO(10)$ model where the gauge symmetry
is broken down to $SU(3)_C\times SU(2)_L \times SU(2)_R
\times U(1)_{B-L}$, to the flipped $SU(5)\times U(1)_X$ symmetry,
or to $SU(3)_C\times SU(2)_L \times U(1)_1 \times U(1)_2$.
However, we cannot improve the SM fermion mass
relations in the $SO(10)$ model if the gauge symmetry
is only broken down to the Pati-Salam $SU(4)_C\times SU(2)_L \times
SU(2)_R$ or the George-Glashow $SU(5)\times
U(1)'$ symmetry. In particular, for the first time
 we generate the realistic SM fermion mass
relation in GUTs by considering the high-dimensional operators in
the K\"ahler potential.

\begin{acknowledgments}

This research was supported in part by the Australian Research
Council under project DP0877916 (CB and FW), by the DOE grant
DE-FG03-95-Er-40917 (TL and DVN), by the Natural Science Foundation
of China under grant No. 10821504 (TL), and by the Mitchell-Heep
Chair in High Energy Physics (TL).

\end{acknowledgments}


\begin{thebibliography}{99} \vspace{-1mm}

\bibitem{Ellis:1990zq}
J.~R.~Ellis, S.~Kelley and D.~V.~Nanopoulos,
Phys.\ Lett.\ B {\bf 249}, 441 (1990);
Phys.\ Lett.\ B {\bf 260}, 131 (1991);
U.~Amaldi, W.~de Boer and H.~Furstenau,
coupling
Phys.\ Lett.\ B {\bf 260}, 447 (1991);
P.~Langacker and M.~X.~Luo,
Phys.\ Rev.\ D {\bf 44}, 817 (1991).


\bibitem{Georgi:1974sy}
H.~Georgi and S.~L.~Glashow,
Phys.\ Rev.\ Lett.\ {\bf 32}, 438 (1974).


\bibitem{so10} H. Georgi, ``Particles And Fields: Williamsburg 1974.
AIP Conference Proceedings No. 23'', Editor C.~E.~Carlson;
H.~Fritzsch and P.~Minkowski,
Annals Phys.\ {\bf 93}, 193 (1975);
H.~Georgi and D.~V.~Nanopoulos,
Nucl.\ Phys.\ B {\bf 155}, 52 (1979).

\bibitem{mSUGRA}
A.~H.~Chamseddine, R.~L.~Arnowitt and P.~Nath,
Phys.\ Rev.\ Lett.\ {\bf 49}, 970 (1982);
R.~Barbieri, S.~Ferrara and C.~A.~Savoy,
Phys.\ Lett.\ B {\bf 119}, 343 (1982);
J.~R.~Ellis, D.~V.~Nanopoulos and K.~Tamvakis,
Phys.\ Lett.\ B {\bf 121}, 123 (1983);
J.~R.~Ellis, J.~S.~Hagelin, D.~V.~Nanopoulos and K.~Tamvakis,
Supergravity,'' Phys.\ Lett.\ B {\bf 125}, 275 (1983);
L.~J.~Hall, J.~D.~Lykken and S.~Weinberg,
Phys.\ Rev.\ D {\bf 27}, 2359 (1983).




\bibitem{Ellis:1984bm}
J.~R.~Ellis, C.~Kounnas and D.~V.~Nanopoulos,
Nucl.\ Phys.\ B {\bf 247}, 373 (1984).



\bibitem{gaugemediation}
M.~Dine, W.~Fischler and M.~Srednicki,
Nucl.\ Phys.\ B {\bf 189}, 575 (1981); S.~Dimopoulos and S.~Raby,
Nucl.\ Phys.\ B {\bf 192}, 353 (1981); M.~Dine and W.~Fischler,
Phys.\ Lett.\ B {\bf 110}, 227 (1982); M. Dine and A. E. Nelson,
Phys. Rev. {\bf D48}, 1277 (1993);
M. Dine, A. E. Nelson and Y. Shirman, Phys. Rev. {\bf D51}, 1362
(1995);
M. Dine, A. E. Nelson, Y. Nir and Y. Shirman, Phys. Rev. {\bf D53},
2658 (1996);
for a review, see G. F. Giudice and R. Rattazzi, Phys. Rept. {\bf
322}, 419 (1999).


\bibitem{anomalymediation}
L.~Randall and R.~Sundrum,
Nucl.\ Phys.\ B {\bf 557}, 79 (1999);
G.~F.~Giudice, M.~A.~Luty, H.~Murayama and R.~Rattazzi,
JHEP {\bf 9812}, 027 (1998).








\bibitem{UVI-AMSB}
I.~Jack and D.~R.~T.~Jones,
Phys.\ Lett.\ B {\bf 482}, 167 (2000);
N.~Arkani-Hamed, D.~E.~Kaplan, H.~Murayama and Y.~Nomura,
JHEP {\bf 0102}, 041 (2001);
R.~Kitano, G.~D.~Kribs and H.~Murayama,
Phys.\ Rev.\ D {\bf 70}, 035001 (2004).


\bibitem{D-AMSB}
A. Pomarol and R. Rattazzi, JHEP {\bf 9905}, 013 (1999);
R. Rattazzi, A. Strumia and J. D. Wells, Nucl. Phys. B {\bf 576}, 3;
N. Okada, Phys. Rev. {\bf D65}, 115009 (2002).




\bibitem{Ellis:1985jn}
J.~R.~Ellis, K.~Enqvist, D.~V.~Nanopoulos and K.~Tamvakis,
Phys.\ Lett.\ B {\bf 155}, 381 (1985).



\bibitem{Hill:1983xh}
C.~T.~Hill,
Phys.\ Lett.\ B {\bf 135}, 47 (1984).


\bibitem{Shafi:1983gz}
Q.~Shafi and C.~Wetterich,
Phys.\ Rev.\ Lett.\ {\bf 52}, 875 (1984).




\bibitem{Drees:1985bx}
M.~Drees,
Phys.\ Lett.\ B {\bf 158}, 409 (1985).




\bibitem{Anderson:1999uia}
G.~Anderson, H.~Baer, C.~h.~Chen and X.~Tata,
Phys.\ Rev.\ D {\bf 61}, 095005 (2000).


\bibitem{Chamoun:2001in}
N.~Chamoun, C.~S.~Huang, C.~Liu and X.~H.~Wu,
Nucl.\ Phys.\ B {\bf 624}, 81 (2002).


\bibitem{Chakrabortty:2008zk}
J.~Chakrabortty and A.~Raychaudhuri,
Phys.\ Lett.\ B {\bf 673}, 57 (2009).


\bibitem{Martin:2009ad}
S.~P.~Martin,
Phys.\ Rev.\ D {\bf 79}, 095019 (2009).

\bibitem{Bhattacharya:2009wv}
S.~Bhattacharya and J.~Chakrabortty,
Phys.\ Rev.\ D {\bf 81}, 015007 (2010).

\bibitem{Feldman:2009zc}
D.~Feldman, Z.~Liu and P.~Nath,
Phys.\ Rev.\ D {\bf 80}, 015007 (2009).


\bibitem{Chamoun:2009nd}
N.~Chamoun, C.~S.~Huang, C.~Liu and X.~H.~Wu,
arXiv:0909.2374 [hep-ph].

\bibitem{stefan}
 Stefan Antusch, Martin Spinrath,
Phys.\ Rev.\ D{\bf 79}, 095004(2009).

\bibitem{india}
 Joydeep Chakrabortty, Amitava Raychaudhuri,
arXiv:1006.1252 [hep-ph].





\bibitem{Vafa:1996xn}
C.~Vafa,
Nucl.\ Phys.\ B {\bf 469}, 403 (1996).


\bibitem{Donagi:2008ca}
R.~Donagi and M.~Wijnholt,
arXiv:0802.2969 [hep-th].

\bibitem{Beasley:2008dc}
C.~Beasley, J.~J.~Heckman and C.~Vafa,
JHEP {\bf 0901}, 058 (2009).

\bibitem{Beasley:2008kw}
C.~Beasley, J.~J.~Heckman and C.~Vafa,
JHEP {\bf 0901}, 059 (2009).


\bibitem{Donagi:2008kj}
R.~Donagi and M.~Wijnholt,
arXiv:0808.2223 [hep-th].




\bibitem{Font:2008id}
A.~Font and L.~E.~Ibanez,
JHEP {\bf 0902}, 016 (2009).





\bibitem{Jiang:2009zza}
J.~Jiang, T.~Li, D.~V.~Nanopoulos and D.~Xie,
Phys.\ Lett.\ B {\bf 677}, 322 (2009).


\bibitem{Blumenhagen:2008aw}
R.~Blumenhagen,
Phys.\ Rev.\ Lett.\ {\bf 102}, 071601 (2009).









\bibitem{Jiang:2009za}
J.~Jiang, T.~Li, D.~V.~Nanopoulos and D.~Xie,
Nucl.\ Phys.\ B {\bf 830}, 195 (2010).






\bibitem{Li:2009cy}
  T.~Li,
  Phys.\ Rev.\  D {\bf 81}, 065018 (2010).



\bibitem{Leontaris:2009wi}
G.~K.~Leontaris and N.~D.~Tracas,
arXiv:0912.1557 [hep-ph].


\bibitem{Li:2010mr}
T.~Li, J.~A.~Maxin and D.~V.~Nanopoulos,
arXiv:1002.1031 [hep-ph].





\bibitem{Li:2010xr}
  T.~Li and D.~V.~Nanopoulos,
  Phys.\ Lett.\  B {\bf 692}, 121 (2010).


\bibitem{Li:2010hi}
T.~Li and D.~V.~Nanopoulos,
arXiv:1005.3798 [hep-ph].



\bibitem{Buras:1977yy}
  A.~J.~Buras, J.~R.~Ellis, M.~K.~Gaillard and D.~V.~Nanopoulos,
  Nucl.\ Phys.\  B {\bf 135}, 66 (1978).


\bibitem{Nanopoulos:1978hh}
  D.~V.~Nanopoulos and D.~A.~Ross,
  Nucl.\ Phys.\  B {\bf 157}, 273 (1979).

\bibitem{Nanopoulos:1982fc}
  D.~V.~Nanopoulos and D.~A.~Ross,
  Phys.\ Lett.\  B {\bf 118}, 99 (1982).




\bibitem{georgi2} H. Georgi and C. Jarlskog, Phys.\ Lett.\ B{\bf 86},297(1979).
\bibitem{gn} H. Georgi and D.V. Nanopoulos, Phys.\ Lett.\ B {\bf 82},392(1979);
Nucl.\ Phys.\ B {\bf 155},52(1979); Nucl.\ Phys.\ B {\bf 159},16
(1979).
\bibitem{ellis} John Ellis, Mary K. Gaillard, Phys.\ Lett.\ B{\bf 88},315(1979).


\bibitem{Nanopoulos:1982zm}
  D.~V.~Nanopoulos and M.~Srednicki,
  Phys.\ Lett.\  B {\bf 124}, 37 (1983).


\bibitem{murayama} J. L. Diaz-Cruz, H. Murayama, A. Pierce,
Phys.\ Rev.\ D{\bf 65}, 075011(2002).




\bibitem{fei}
C.~Balazs, T.~Li, D.~V.~Nanopoulos and F.~Wang,
  JHEP {\bf 1009}, 003 (2010).






\bibitem{smbarr} S. M. Barr,
Phys.\ Lett.\ B {\bf 112}, 219 (1982).


\bibitem{dimitri}
J.~P.~Derendinger, J.~E.~Kim and D.~V.~Nanopoulos,
Phys.\ Lett.\ B {\bf 139}, 170 (1984).

\bibitem{AEHN-0}
I.~Antoniadis, J.~R.~Ellis, J.~S.~Hagelin and D.~V.~Nanopoulos,
Phys.\ Lett.\ B {\bf 194}, 231 (1987).



\bibitem{Slansky:1981yr}
R.~Slansky,
Phys.\ Rept.\ {\bf 79}, 1 (1981).

\bibitem{ross} Graham Ross, Mario Serna, Phys.\ Lett.\ B{\bf 664},97(2008).




\bibitem{He:1990jw}
X.~G.~He and S.~Meljanac,
Phys.\ Rev.\ D {\bf 41}, 1620 (1990).






\end{thebibliography}
\end{document}